\documentclass[a4paper,12pt]{article}

\usepackage{cmap}     
\usepackage{amsmath,amssymb}
\usepackage[utf8]{inputenc}
\usepackage[T2A]{fontenc}  
\usepackage{graphicx}    
\usepackage[margin=1in]{geometry}
\usepackage{fancyhdr}    
\usepackage{setspace}
\usepackage{wrapfig}
\usepackage{subfig}
\usepackage{listings}
\usepackage{color}
\usepackage{setspace}
\usepackage{textcomp}
\usepackage{float}
\usepackage{listings}
\usepackage{amsthm}
\usepackage{algorithm}
\usepackage{algpseudocode}
\usepackage{afterpage}
\usepackage[toc,page]{appendix}
\usepackage{natbib}
\usepackage{bbm}
\usepackage{xcolor}
\usepackage{footnote}
\makesavenoteenv{tabular}
\usepackage{enumerate}

\newtheorem{theorem}{Theorem}

\newtheorem{proposition}{Proposition}

\DeclareMathOperator{\diag}{diag}

\DeclareMathOperator{\inter}{int}

\DeclareMathOperator{\errL}{err}

\begin{document}


\title{Transition probability of Brownian motion in the octant and its application to default modeling} 
\author{Vadim Kaushansky\thanks{The first author gratefully acknowledges support from the Economic and Social Research Council and Bank of America Merrill Lynch}
\footnote{\emph{Corresponding author}, Mathematical Institute \& Oxford-Man Institute, University of Oxford, UK, E-mail: vadim.kaushansky@maths.ox.ac.uk},  Alexander Lipton\footnote{Massachusetts Institute of Technology, Connection Science, Cambridge, MA, USA and École Polytechnique Fédérale de Lausanne, Switzerland, E-mail: alexlipt@mit.edu}, Christoph Reisinger\footnote{Mathematical Institute  \& Oxford-Man Institute, University of Oxford, Andrew Wiles Building, Woodstock Road, Oxford, OX2 6GG, UK, E-mail: christoph.reisinger@maths.ox.ac.uk}}    
\date{}
   
\maketitle 
\begin{abstract}
We derive a semi-analytical formula for the transition probability of three-dim\-ensional Brownian motion in the positive octant with absorption at the boundaries. 
Separation of variables in spherical coordinates leads to an eigenvalue problem for the resulting boundary value problem in the two angular components.
The main theoretical result is a solution to the original problem expressed as an expansion into special functions and an eigenvalue which has to be chosen to allow a matching of the boundary condition. We discuss and test several computational methods to solve a finite-dimensional approximation to this nonlinear eigenvalue problem. Finally, we apply our results to the computation of default probabilities and credit valuation adjustments in a structural credit model with mutual liabilities.
\end{abstract}

\medskip

\section{Introduction}
The transition probability of multi-dimensional Brownian motion in $\mathbb{R}^n_{+}, n \ge 1,$ with absorption at the boundary, occurs in different areas of applied mathematics. We are particularly motivated by mathematical finance. For example, it appears in counterparty credit risk modeling (\cite{LiptonSav}, \cite{LiptonSepp2009}, \cite{LiptonItkin2015}, \cite{Zhou2001a}), market microstructure (\cite{cont2012order}, \cite{Lipton2014}), and exotic option pricing (\cite{escobar2014barrier} for barrier options on three assets with special correlation structure; \cite{He1998} for lookback options whose payoff depends on the minima, maxima, and endpoint of the asset price process; \cite{Lipton2001}).

The one-dimensional case is classical and is solved analytically by the reflection principle, or, in PDE terms, by the method of images. The two-dimensional problem can be solved analytically by the method of images or separation of variables; see \cite{LiptonSepp2009} for details. The three-dimensional problem is much more difficult than the one- and two-dimensional ones.  For several particular cases of the correlation matrix, \cite{escobar2013three} offered solutions to the three-dimensional problem by the method of images. Unfortunately, it cannot be extended to general correlation matrices. A semi-analytical solution was proposed in \cite{LiptonSav}, where a separation of variables technique is used and the problem is split into radial and angular PDEs. The radial PDE is solved analytically, while the angular PDE is solved by finite elements.

In this paper, we use the separation of variables technique and solve the angular PDE semi-analytically, and thus provide a ``fully semi-analytical'' representation of the Green's function.
By this we mean that the solution is expressed as a (double) infinite series of terms involving special functions (hypergeometric and Bessel). We also require the solution of a countably infinite-dimensional non-linear eigenvalue problem. In practice, the sums can be truncated to a relatively small number of terms (low tens) without significant loss of accuracy, as evidenced by our numerical tests, and similarly for the dimension of the non-linear eigenvalue problem. 

There are a number of published methods available for solving (finite-dimensional) non-linear eigenvalue problems. A detailed review can be found in \cite{nonlineig}. Most methods use a linearization and solution of a corresponding linear eigenvalue problem. As an example of this type of methods, we consider a method based on Chebyshev interpolation (\cite{ApproxTheory}). We also study another recently developed approach based on computing contour integrals (\cite{Beyn}). Both methods are easy to implement and have good properties for our specific problem.

By this semi-analytical approach, we can compute the transition density and its derivatives to high accuracy much more quickly than by numerical solution of the original three-dimensional parabolic PDE. We demonstrate this by comparison with a finite difference approximation obtained with the Hundsdorfer-Verwer scheme (\cite{HV}).

To our knowledge, this is the first published expression for the transition probability of general correlated Brownian motion with absorbing boundaries on the octant.
Moreover, we consider a slightly more general formulation compared to other published special cases which allows a constant drift. 

In the context of credit risk applications, we show how to compute the joint survival probability of firms in a Black-Cox setting semi-analytically using Gegenbauer polynomial expansion, and give expressions for
credit- and debt-valuation adjustments in terms of integrals over (analytically available) derivatives of the transition probability.

The rest of the paper is organized as follows. In Section 2, we formulate the problem as PDE, show how to eliminate the drift and correlation terms, and transform the equation to  spherical coordinates. In Section 3, we use separation of variables to obtain radial and angular PDEs, solve both equations, and write the final expression for the Green's function
in terms of unknown non-linear eigenvalues and eigenvectors. In Section 4, we discuss various methods to solve the nonlinear eigenvalue problem. In Section 5, we perform numerical tests. In Section 6, we study applications of the three-dimensional transition probability to a default model with mutual liabilities. In Section 7, we conclude.

\section{Problem formulation and preliminaries}
We consider a three-dimensional standard Brownian motion with constant drift $(\mu_i)_{1\le i\le 3}$,
\begin{eqnarray*}
	\label{X_t_eq}
	d X_s^i = \mu_i \, d s +  d W_s^i, \qquad
	X_t^i = x_i, \qquad 1\le i\le 3,
\end{eqnarray*}
with infinitesimal generator
\begin{equation*}
	\mathcal{L}  = \frac{1}{2} \sum_{i, j = 1}^3  \rho_{ij} \frac{\partial^2 }{\partial x_i \partial x_j} + \sum_{i = 1}^3  \mu_i \frac{\partial }{\partial x_i},
\end{equation*}
where $(\rho_{ij})_{1 \le i, j \le 3}$ is a symmetric positive definite matrix with $\rho_{ii} = 1$, $1\le i\le 3$, the correlation matrix of the three-dimensional standard Brownian motion
$W$.

Now consider the stopped process $X_{s\wedge \tau}$, with $\tau = \inf \{u : X_u \notin \mathbb{R}^3_{+} \}$ the exit time of the positive octant.
Then the distribution of $X_{s\wedge \tau}$ has a singular component at the boundary where one of the coordinates is 0, corresponding to the absorbed mass, and
a continuous density in the interior, which goes to zero at the boundaries.
Let $p(t', x' | t, x)$ denote this continuous component of the distribution 
evaluated at $x' \in \mathbb{R}^3_{+}$ at time $t'\ge t$.
For a fixed pair $(t, x)$, 
it satisfies the forward Kolmogorov equation
\begin{equation}
	\label{adjoint_eq}
	\frac{\partial p}{\partial t'}(t', x' | t, x) = \mathcal{L}^* p(t', x' | t, x),
\end{equation}
with initial condition $p(t, x' | t, x) = \delta(x' - x)$,
boundary condition $p(t', x' | t, x) = 0$ for $x_1' \cdot x_2' \cdot x_3' = 0$, $t' \ge t$,
 and where $\mathcal{L}^*$ is the adjoint operator of $\mathcal{L}$,
\begin{equation*}
	\mathcal{L}^* = \frac{1}{2} \sum_{i, j = 1}^3 \rho_{ij} \frac{\partial^2}{\partial x'_i \partial x'_j}  - \sum_{i = 1}^3  \mu_i \frac{\partial}{\partial x'_i} .
\end{equation*}

In PDE theory, the transition probability can be interpreted as Green's function, that is the fundamental solution to a suitable initial(-boundary) value problem which we will specify later. Thus, by time homogeneity, we use the notation $G(t'-t, x', x) = p(t', x'| t, x)$. 

For convenience, we will denote the state variables occasionally by $(x, y, z) \in \mathbb{R}^3_{+}$.

To eliminate the cross-derivative terms, we use the change of variables $(x,y,z) \rightarrow (\alpha, \beta, \gamma)$ as in \cite{Lipton2014} and \cite{LiptonSav}; see also
Appendix \ref{cross_deriv}. 
To eliminate the drift term, we further use the transformation
\begin{eqnarray}
\nonumber
	\tilde{G}(t'-t,  \alpha', \beta', \gamma', \alpha, \beta, \gamma) &\!\!=\!\!& \exp\Big(-\xi_{\alpha}(\alpha-\alpha') - \xi_{\beta}(\beta-\beta') - \xi_{\gamma}(\gamma - \gamma')- \\
	&& \qquad \frac{\xi_{\alpha}^2+\xi_{\beta}^2+\xi_{\gamma}^2}{2}(t-t')\Big)	\times 
	G(t'-t, \alpha', \beta', \gamma', \alpha, \beta, \gamma)
	\label{non_drift}
\end{eqnarray}
(see also Appendix \ref{cross_deriv}). We omit tilde in the following for simplicity of notation.

Finally, we rewrite the equation in spherical coordinates $(r, \varphi, \theta)$ to take advantage of the wedge-shape of the domain, where $r \in (0,\infty)$,
$\varphi \in (0,\bar{\omega})$ with  $\bar{\omega} = \arccos(-\rho_{xy})$, and $\theta \in (0, \Theta(\varphi))$, where $\Theta(\varphi)$ can be implicitly defined as
\begin{eqnarray}
\varphi(\omega) &=& \arccos \left(\frac{1 - \rho_{xy} \omega}{\sqrt{1 - 2 \rho_{xy} \omega + \omega^2}} \right), \label{varphi} \\
\Theta(\omega) &=& \arccos \left(- \frac{\rho_{yz}  - \rho_{xz} \rho_{xy} + \omega (\rho_{xz} - \rho_{yz} \rho_{xy})}{\sqrt{\bar{\rho}_{xy} (\bar{\rho}_{xz}^2 - 2 \omega (\rho_{xy} - \rho_{xz} \rho_{yz} ) + \omega^2 \bar{\rho}_{yz}^2 )}}  \right), \label{theta}
\end{eqnarray}
where $\bar{\rho} = \sqrt{1 - \rho^2}$ for $\rho \in \{{\rho}_{xy}, {\rho}_{yz}\}$.

The final initial-boundary value problem for the Green's function has the form
\begin{equation}
	\left\{
	\label{3d_eq_spherical}	
		\begin{aligned}
			& \frac{\partial G}{\partial t'} - \frac{1}{2} \left[\frac{1}{r'} \frac{\partial^2}{\partial r'^2} (r' G)  + \frac{1}{r'^2} \left( \frac{1}{\sin^2{\theta'}} G_{\varphi' \varphi'} + \frac{1}{\sin{\theta'}} \frac{\partial}{\partial \theta'}(\sin{\theta'} G_{\theta'} \right) \right] = 0, \vspace{0 cm} \\
			& G(0, r', \varphi', \theta', r, \varphi, \theta) = \frac{1}{r^2 \sin{\theta}} \delta(r' - r) \delta(\varphi' - \varphi) \delta(\theta'-\theta),  \\ \vspace{0 cm}
			& G(t'-t, r', 0, \theta', r, \varphi, \theta) = 0, \quad\qquad G(t'-t, r', \bar{\omega}, \theta', r, \varphi, \theta) = 0, \\
			& G(t'-t, r', \varphi', 0, r, \varphi, \theta) = 0, \quad\qquad G(t'-t, r', \varphi',  \Theta(\varphi'), r, \varphi, \theta) = 0, \\			
			& G(t'-t, 0, \varphi', \theta', r, \varphi, \theta)  = 0, \quad\qquad  G(t'-t, r', \varphi', \theta' , r, \varphi, \theta) \underset{r' \ \to +\infty}{\longrightarrow} 0.
		\end{aligned}
	\right.
\end{equation}

\section{Semi-analytical solution by expansion}
\label{sec:sepvar}

In this section, we show how the solution to
(\ref{3d_eq_spherical}) can be given via an eigenvalue expansion. First, we apply separation of variables
\begin{equation*}
	G(t, r', \varphi', \theta', r, \varphi, \theta) = g(t, r', r) \Psi(\varphi', \theta', \varphi, \theta),
\end{equation*}
then (\ref{3d_eq_spherical}) decomposes into two separate (initial-)boundary value problems
\begin{equation}
	\label{radial_pde}
	\left\{
	\begin{aligned}
		& g_{t'} = \frac{1}{2} \left(\frac{1}{r'} \frac{\partial^2}{\partial r'^2} (r' g) - \frac{\Lambda^2}{r'^2} g \right), \\
		& g(0, r', r) = \frac{1}{r} \delta(r' - r), \\
		& g(t'-t, 0, r) = 0, \quad g(t'-t, r', r) \underset{r' \ \to +\infty}{\longrightarrow} 0,
	\end{aligned}
	\right.
\end{equation}
and
\begin{equation}
	\label{angular_pde}
	\left\{
	\begin{aligned}
		& \frac{1}{\sin^2{\theta'}} \Psi_{\varphi' \varphi'} + \frac{1}{\sin{\theta'}} \frac{\partial}{\partial \theta'} (\sin{\theta'} \Psi_{\theta'}) = -\Lambda^2 \Psi, \\
		& \Psi(0, \theta', \varphi, \theta) = 0, \Psi(\bar{\omega}, \theta', \varphi, \theta) = 0, \Psi(\varphi', 0, \varphi, \theta) = 0, \Psi(\varphi', \Theta(\varphi'), \varphi, \theta) = 0,
	\end{aligned}
	\right.
\end{equation}
where $\Lambda^2 \in \mathbb{R}_{+}$ is an eigenvalue.
The eigenvalue is real and positive because the differential operator on the left-hand side of (\ref{angular_pde}) is symmetric and positive definite with respect to a certain weighted inner product, as is seen from
the variational formulation given in equation (40) in \cite{LiptonSav}.

{\color{black} We note an asymmetry between (\ref{radial_pde}), which contains the time variable and hence an initial condition is imposed, and
(\ref{angular_pde}), which is stationary and hence does not match any initial condition.
The initial condition of the solution in (\ref{3d_eq_spherical}) will eventually be ensured by superposition in Section \ref{subsec:final}.
Since \eqref{angular_pde} does not depend on $(\varphi, \theta)$ here (due to the ignorance of the initial condition), we further write $\Psi(\varphi', \theta', \varphi, \theta) = \Psi(\varphi', \theta')$ for simplicity.}

	The PDE (\ref{radial_pde}) can be solved analytically, with solution
	\begin{equation*}
		g(t, r', r) = \frac{e^{-\frac{r^2+r'^2}{2 t}}}{t \sqrt{r r'}} I_{\sqrt{\Lambda^2 + \frac{1}{4}}} \left(\frac{r r'}{t} \right),
	\end{equation*}
	where  $I_a(b)$ is the modified Bessel function of the second kind.
	

The eigenvalue problem for the angular PDE (\ref{angular_pde})  was solved numerically in \cite{LiptonSav} by a finite element method.
In the following, we consider an alternative, semi-analytical approach.

\subsection{Semi-analytical method for angular PDE}

We begin by further simplifying the problem by the extra transformation (as in \cite{Lipton2014} for the stationary case)
\begin{equation}
	\label{zeta_transform}
	\zeta' = \ln{\tan{\theta'/2}},
\end{equation} 
which changes the domain to the semi-infinite strip $-\infty < \zeta \le Z(\varphi) =  \ln{\tan{\Theta(\varphi)/2}}$ and the eigenvalue problem to
\begin{equation}
	\label{angular_pde_mod}
	\left\{
	\begin{aligned}
		& \Psi_{\varphi' \varphi'} +  \Psi_{\zeta' \zeta'} = -4\Lambda^2 \frac{e^{2 \zeta'}}{(1 + e^{2 \zeta'})^2} \Psi, \\
		& \Psi(0, \zeta') =  \Psi(\bar{\omega}, \zeta') = \lim_{\zeta'\rightarrow -\infty} \Psi(\varphi', \zeta') = \Psi(\varphi', Z(\varphi')) = 0.
	\end{aligned}
	\right.
\end{equation}

Denote $\lambda = \frac{1}{2} + \sqrt{\frac{1}{4} + \Lambda^2}$. Then, (\ref{angular_pde_mod}) becomes
\begin{equation*}
	 \Psi_{\varphi' \varphi'} +  \Psi_{\zeta' \zeta'} = -\frac{\lambda (\lambda - 1)}{\cosh^2(\zeta')} \Psi. \\
\end{equation*}

We shall find solutions in the form $\Psi(\varphi', \zeta') = \Phi(\varphi') \cdot \Upsilon(\zeta')$. Thereby, we can split (\ref{angular_pde_mod}) into the system

\begin{align}
		\label{phi_eq}
		 & \Phi'' = -k^2 \Phi, \\ 
		 \label{ups_eq1}
		 & \Upsilon'' = \left(k^2 - \frac{\lambda (\lambda - 1)}{\cosh^2(\zeta')}  \right) \Upsilon,
\end{align}
where $k^2 \in \mathbb{R}_{+}$ is an eigenvalue and with boundary conditions
\begin{equation}
	\label{bound_conditions}
	\Phi(0) = 0,  \quad \Phi(\bar{\omega}) = 0, \quad \Upsilon(\zeta') \underset{\zeta\ \to -\infty}{\longrightarrow} 0,  \quad \Psi(\varphi', Z(\varphi')) = 0.
\end{equation}
Equation (\ref{phi_eq}) with the first and second boundary conditions in (\ref{bound_conditions}) can be solved  as
\begin{equation*}
	\Phi_n(\varphi') = c_n \sin{(k_n \varphi')},
\end{equation*}
where $k_n = \pi n / \bar{\omega}$, while
(\ref{ups_eq1}) is the Schrödinger equation with Pöschl–Teller potential. Solving it with the third boundary condition in (\ref{bound_conditions}), we get (see Appendix \ref{AppendixComputations})
\begin{equation*}
	\Upsilon_n(\zeta') = c_n e^{k_n \zeta'} {}_{2}F_1 \left(\lambda, 1 - \lambda, 1 + k_n, \frac{e^{2 \zeta'}}{1+e^{2 \zeta'}} \right),
\end{equation*}
where ${}_2F_1(a, b, c, z)$ is the Gaussian hypergeometric function.
Thus, we find $\Psi(\varphi', \zeta')$ in the form
\begin{equation}
	\label{Psi_expression}
	\Psi(\varphi', \zeta') = \sum_{n = 1}^{\infty} c_n \sin{(k_n \varphi')} e^{k_n \zeta'} {}_{2}F_1 \left(\lambda, 1 - \lambda, 1 + k_n, \frac{e^{2 \zeta'}}{1+e^{2 \zeta'}} \right).
\end{equation}
As an aside, we note that $ {}_{2}F_1 \left(\lambda, 1 - \lambda, 1 + k_n, \frac{e^{2 \zeta'}}{1+e^{2 \zeta'}} \right) = 1$ for $\lambda = 0$, which gives
\begin{equation*}
	\label{Psi_lambda0}
	\Psi_0(\varphi', \zeta')  = \sum_{n = 1}^{\infty} c_n \sin{(k_n \varphi')} e^{k_n \zeta'}
\end{equation*}
and coincides with the expression in \cite{Lipton2014} for the stationary problem. 

To find the coefficients $c_n$ and eigenvalues $\lambda$, we use the fourth condition in (\ref{bound_conditions}),
\begin{equation}
	\label{fourth_bc}
	\sum_{n = 1}^{\infty} c_n \sin{(k_n \varphi)} e^{k_n Z(\varphi')} {}_{2}F_1 \left(\lambda, 1 - \lambda, 1 + k_n, \frac{e^{2 Z(\varphi')}}{1+e^{2 Z(\varphi')}} \right) = 0
\end{equation}
for all $\varphi' \in [0, \bar{\omega}]$.

We introduce the integrals
\begin{eqnarray}
	\label{J_mn}
	T_{mn}(\lambda) &=&
	\langle f_m, f_n
	\rangle \; = \;
	\int \limits_0^{\bar{\omega}}
	f_m(\varphi') f_n(\varphi') \, d \varphi', \\
f_n(\varphi') &=& \sin{(k_n \varphi')} e^{ k_n Z(\varphi')} {}_{2}F_1 \left(\lambda, 1 - \lambda, 1 + k_n, \frac{e^{2 Z(\varphi')}}{1+e^{2 Z(\varphi')}} \right),
\nonumber
\end{eqnarray}
and note that $T_{mn}$ is a function of $\lambda$ only. Then, applying (\ref{fourth_bc}), we get
\begin{equation}
	\label{Jc_eq_inf}
	\sum_{n = 1}^{\infty} T_{mn}(\lambda) c_n = 0, \quad \forall m \ge 1.
\end{equation}
We defer the solution of (\ref{Jc_eq_inf}) for $\lambda$ and $(c_n)_{n\ge 1}$ to Section \ref{NEP}.

\subsection{Final expression for Green's function}
\label{subsec:final}
Now we can write the expression for the Green's function using the eigenvalue expansion
\begin{eqnarray}
	G(\tau, r', \varphi', \zeta', r, \varphi, \zeta) &=& \sum_{l = 1}^{\infty} a_l g_l(\tau, r') \Psi_l(\varphi', \zeta') \nonumber \\
	&=&  \sum_{l = 1}^{\infty} a_l g_l(\tau, r') \sum_{n = 1}^{\infty} c_n^l \Phi_n(\varphi') \Upsilon_n^l(\zeta') \nonumber \\
	 &\hspace{-8 cm}=&\hspace{-4.1 cm}  \frac{e^{-\frac{r'^2+r^2}{2 \tau}}}{\tau \sqrt{r' r}}\sum_{l = 1}^{\infty} a_l  I_{\lambda_l} \! \left(\frac{r' r}{\tau}\right) \! \times \! 
	 \sum_{n = 1}^{\infty} c_n^l \sin{\left(k_n \varphi' \right)} e^{k_n \zeta'} {}_{2}F_1 \! \left(\lambda_l, 1 - \lambda_l, 1 + k_n, \frac{e^{2 \zeta'}}{1+e^{2 \zeta'}} \right)\!,\;\;\;
	 \label{greens_pre_final}
\end{eqnarray}
where the (countably many) eigenvalues $\lambda_l$ and coefficients $c_n^l$ can be determined as the solution of the nonlinear eigenvalue problem (\ref{Jc_eq_inf}) and $k_n = \frac{\pi n}{\bar{\omega}}$.

The coefficients $a_l$ can be determined by imposing the initial condition
\begin{equation*}
	G(0, r', \varphi', \theta', r, \varphi, \theta) = \frac{1}{r^2 \sin{\theta}} \delta(r' - r) \delta(\varphi' - \varphi) \delta(\theta' - \theta).
\end{equation*}
We have already ensured the initial condition for the radial part, thus we need
\begin{equation}
	\label{greens_initial}
	\sum_{l = 1}^{\infty} a_l \Psi_l(\varphi', \theta') = \frac{1}{\sin{\theta}} \delta(\varphi' - \varphi) \delta(\theta' - \theta).
\end{equation}

From the weak formulation of the angular PDE it is easy to see that the eigenvectors are orthogonal in the scalar product weighted by $\sin{\theta'}$ (\cite{LiptonSav}).
Multiplying (\ref{greens_initial}) by $\Psi_m(\varphi', \theta') \sin{\theta'}$ and integrating over the whole domain,
\begin{equation*}
	a_l = \iint \limits_{\Omega} \frac{1}{\sin{\theta}} \Psi_l(\varphi', \theta') \sin{\theta'} \delta(\varphi' - \varphi) \delta(\theta' - \theta) \, d \varphi' d \theta' = \Psi_l(\varphi, \theta).
\end{equation*}
Substituting the expression for $a_l$ from the last equation into (\ref{greens_pre_final}), 
expanding the expression for $\Psi_l(\varphi', \zeta')$ and $\Psi_l(\varphi, \zeta)$, we get 
\begin{multline*}
		G(\tau, r', \varphi', \zeta', r, \varphi, \zeta)  =  \frac{e^{-\frac{r'^2+r^2}{2 \tau}}}{\tau \sqrt{r' r}}\sum_{l = 1}^{\infty} I_{\lambda_l} \left(\frac{r' r}{\tau}\right) \times \\
		\times \left(\sum_{n = 1}^{\infty} c_n^l \sin{\left(k_n \varphi' \right)} e^{k_n \zeta'} {}_{2}F_1 \left(\lambda_l, 1 - \lambda_l, 1 + k_n, \frac{e^{2 \zeta'}}{1+e^{2 \zeta'}} \right)\right) \times \\
		\times  \left(\sum_{n = 1}^{\infty} c_n^l \sin{\left(k_n \varphi \right)} e^{k_n \zeta} {}_{2}F_1 \left(\lambda_l, 1 - \lambda_l, 1 + k_n, \frac{e^{2 \zeta}}{1+e^{2 \zeta}} \right)\right). 
\end{multline*}
Returning to the variable $\theta$, we have
\begin{multline}
	\label{greens_final}
		G(\tau, r', \varphi', \theta', r, \varphi, \theta)  =  \frac{e^{-\frac{r'^2+r^2}{2 \tau}}}{\tau \sqrt{r' r}}\sum_{l = 1}^{\infty} I_{\lambda_l} \left(\frac{r' r}{\tau}\right) \times \\
		\times \left(\sum_{n = 1}^{\infty} c_n^l \sin{\left(k_n\varphi' \right)}  \tan^{k_n}\left(\frac{\theta'}{2}\right) {}_{2}F_1 \left(\lambda_l, 1 - \lambda_l, 1 + k_n, \sin^2 \frac{\theta'}{2} \right)\right) \times \\
		\times  \left(\sum_{n = 1}^{\infty} c_n^l \sin{\left(k_n \varphi \right)} \tan^{k_n}\left(\frac{\theta}{2}\right) {}_{2}F_1 \left(\lambda_l, 1 - \lambda_l, 1 + k_n, \sin^2 \frac{\theta}{2} \right)\right). 
\end{multline}

\section{Solution of the nonlinear eigenvalue problem}
\label{NEP}

In this section, we first give properties of the eigenvalue problem (\ref{Jc_eq_inf}).
Then, we describe several methods to solve nonlinear eigenvalue problems and outline how the methods are applied in the tests.

\subsection{Truncation and the smallest (linear) eigenvalue of $T(\lambda)$}
\label{subsec:cheby1}

If we truncate the sum (\ref{Jc_eq_inf}) after $N$ terms, the resulting equation 
can be rewritten in matrix form
\begin{equation}
	T(\lambda) \cdot c = 0,
	\label{Jc_eq_0}
\end{equation}
where $T(\lambda) \in \mathbb{R}^{N \times N}, c \in \mathbb{R}^{N}$.  

\begin{figure}[H]
	\begin{center}
				\subfloat[]{\includegraphics[width=0.5\textwidth]{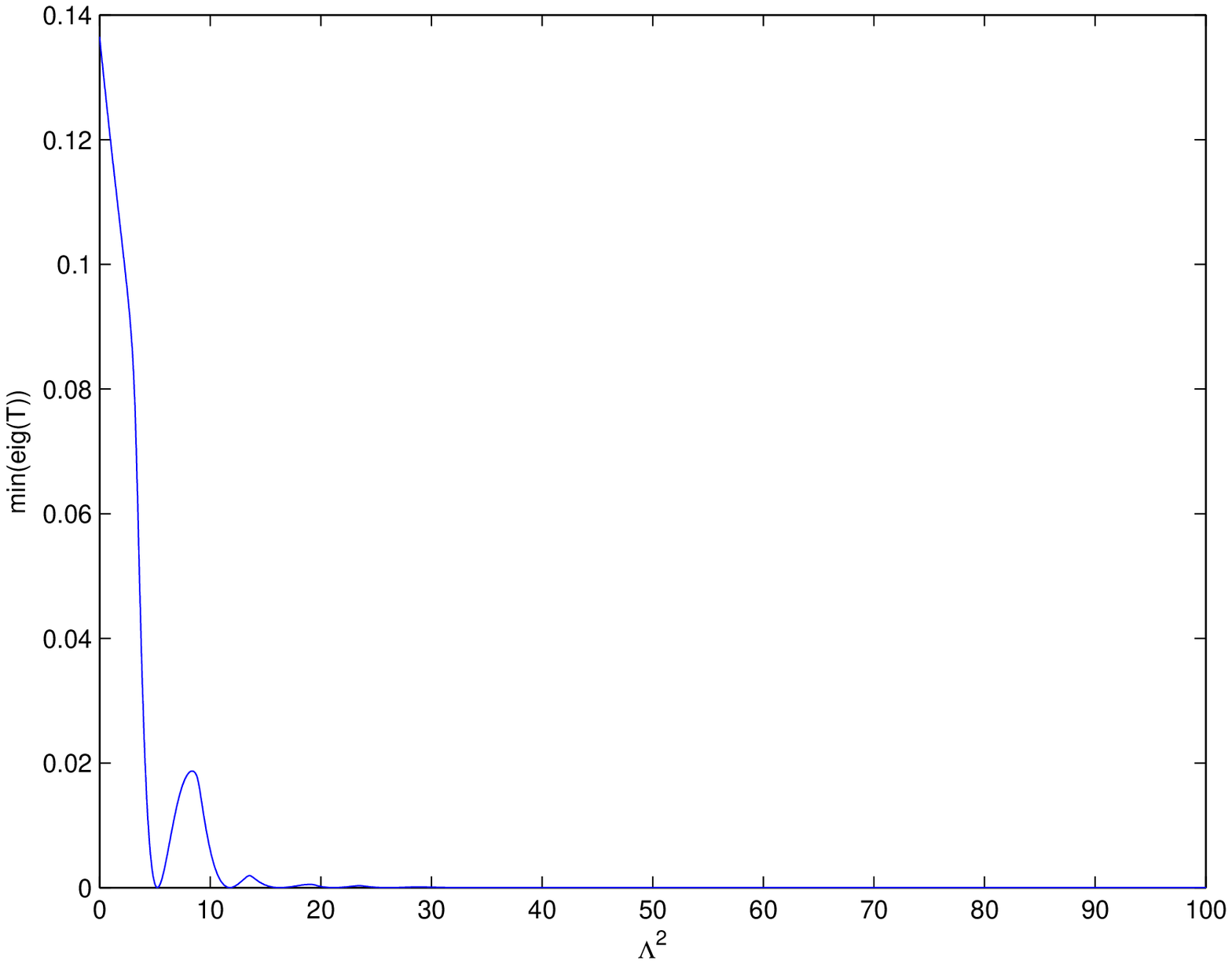}}
				\subfloat[]{\includegraphics[width=0.5\textwidth]{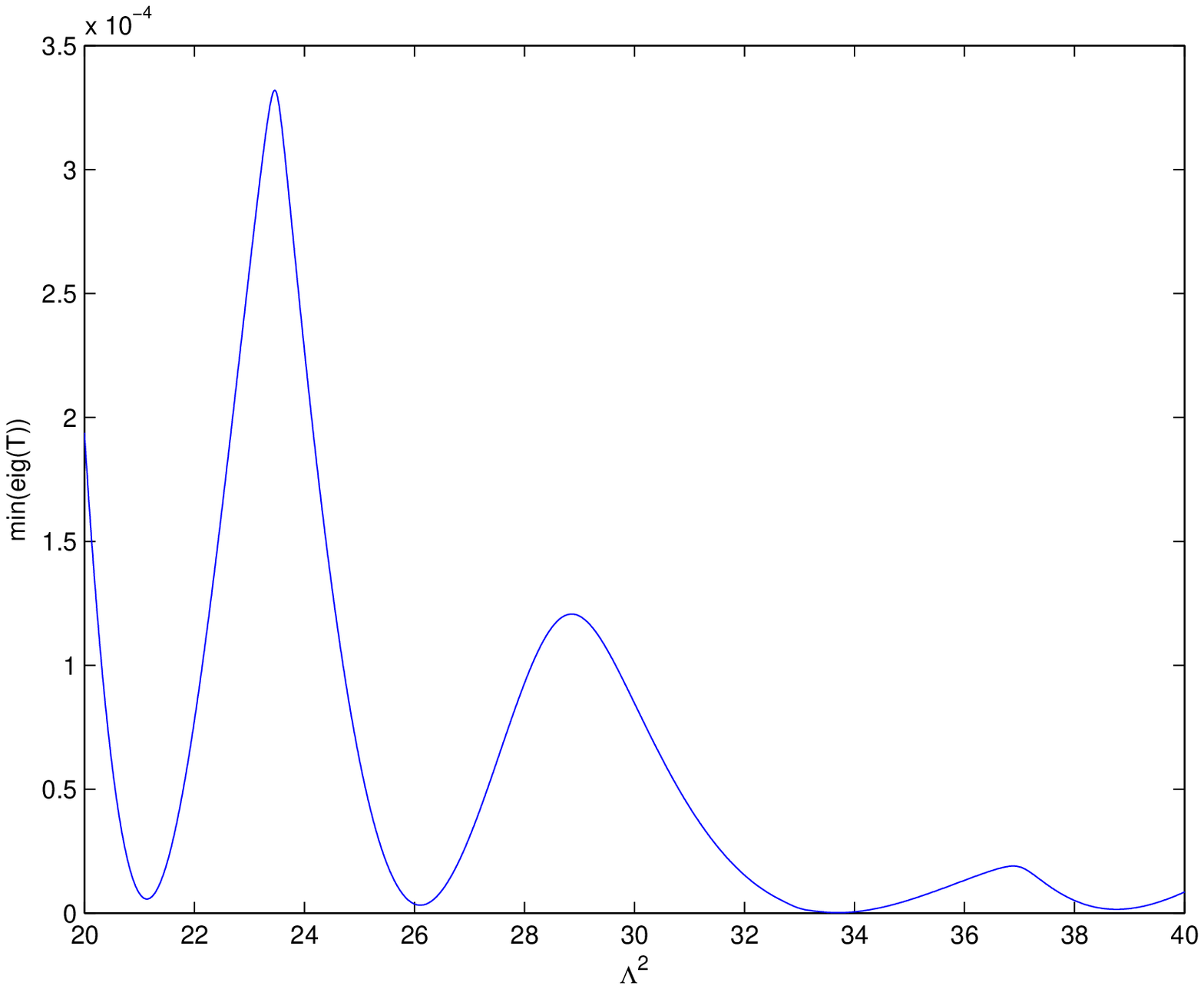}}\\
	\end{center}		
	\vspace{-20pt}
	\caption{Minimum eigenvalue of matrix $T$: $\mu_{min}(T(\lambda))$ for $\rho_{xy} = 0.8, \rho_{xz} = 0.2, \rho_{yz} = 0.5$ and $N = 15$. Note the different plot ranges in (a) and (b).} 
	\label{chebfun}
\end{figure}

Equation (\ref{Jc_eq_0}) is a nonlinear eigenvalue problem for a symmetric positive semi-definite matrix
(as the matrix $T$ is a Gram matrix).
	Eigenvalues $\lambda$ of this truncated problem have to satisfy the equation
	\begin{equation}
		\label{eigval_eq0}
		\mu_{min}(T(\lambda)) = 0,
	\end{equation}
	where $\mu_{min}$ is the minimum eigenvalue of the positive semi-definite matrix $T(\lambda)$.
	Due to the truncation, however, the problem (\ref{Jc_eq_0}) generally does not have real-valued solutions for $\lambda$. 
	Indeed, numerical tests suggest that $T$ is typically strictly positive definite for all $\lambda$ and then a non-zero solution for $c$ does not exist.
	Figure \ref{chebfun} illustrates the dependence of the smallest eigenvalue $\mu_{min}(T(\lambda))$ of $T(\lambda)$ on $\lambda$.

	In order to find an approximation to the original problem, we need to ``regularize'' it in the sense that there exists a solution of the truncated problem that is close to the solution of the infinite problem when the number of terms increases.

	A first simple approach is based on the minima of $\mu_{min}$ (instead of its roots). 
	Here, we use Chebyshev polynomials (\cite{Trefethen}) to approximate all local minima of the left-hand side of (\ref{eigval_eq0}), 
and use these as approximation to the nonlinear eigenvalues.

	More sophisticated and efficient methods by considering complex nonlinear eigenvalues, based on
	Chebyshev interpolation of $T(\lambda)$ directly and on contour integration, respectively,
	are given in Sections \ref{subsec:cheby2} and \ref{subsec:contour} below.\footnote{
	We avoid using $\int\limits_0^{\bar{\omega}} f(\varphi) \bar{g}(\varphi) \, d\varphi$ instead of $\langle \cdot,\cdot\rangle$ in (\ref{J_mn}) due to the positive definiteness.}

\subsection{Chebyshev polynomial interpolation of $T(\lambda)$}
\label{subsec:cheby2}

In this approach, we approximate $T(\lambda)$ for complex $\lambda$ using Chebyshev polynomials,
\begin{equation}
	\label{pol_eigval}
	T(\lambda) \approx \widehat{T}(\lambda) \equiv \sum_{j = 0}^{m} A_j \tilde{T}_j(\lambda),
\end{equation}
where $\{\tilde{T}_j, 0\le j\le m \}$ is a (scalar) Chebyshev polynomial basis of degree $m$ and  $A_j \in \mathbb{R}^{N \times N}$ are the corresponding coefficients.

The following result is a modification of Theorem 18.1 in \cite{ApproxTheory} (see, for example, \cite{nakatsukasa2015computing}, Section 7.2).
\begin{theorem}
The solutions of the polynomial eigenvalue problem for $\widehat{T}$ in \eqref{pol_eigval} are the eigenvalues of the linear matrix pencil (of dimension $mN\times mN$)
\begin{equation}
	 \lambda \left[
			\begin{matrix}
				A_m &  &  \\
				& I_N &\\
				& & \ddots & \ddots & \ddots \\
				& & & & & \\
				& & & &   & I_N \\
			\end{matrix}
		\right]
		- \frac{1}{2} \left[ \begin{matrix}
				-A_{m-1} & I_{\!N}\!-\!A_{m-2} & -A_{m-3} & \cdots & A_0 \\		
				I_N & 0 & I_N\\
				& \ddots & \ddots & \ddots \\
				& & I_N & 0 & I_N \\
				& & &   2 I_N& 0 \\
			\end{matrix}
		\right],
		\label{pencil}
\end{equation}
(i.e., all $\lambda$ such that the above matrix is singular), where entries not displayed are zero.
\end{theorem}

Thus, by finding the eigenvalues of the linear matrix pencil we obtain the solution of a polynomial eigenvalue problem, which is an approximation to the nonlinear eigenvalue problem \eqref{Jc_eq_0}. 

\subsection{Contour integral method}
\label{subsec:contour}

The following method due to \cite{Beyn} finds nonlinear complex eigenvalues within a given contour $\Gamma$.
We give the main steps in Algorithm \ref{algo2} and subsequently discuss the choice of contour.
\begin{algorithm}
	\caption{Contour integral algorithm for nonlinear eigenvalue problem}
	\begin{algorithmic}[1]
		\State{ $l = 1$}
		\State{Choose $\hat{V} \in \mathbb{C}^{N\times l}$ an arbitrary matrix.}
		\State{Compute $A_0 =  \frac{1}{2 \pi i} \int_{\Gamma} T(z)^{-1} \hat{V} \, d z$ and $A_1 = \frac{1}{2 \pi i} \int_{\Gamma} z T(z)^{-1} \hat{V} \, d z$ numerically.}
		\State{Compute SVD $A_0 = V \Sigma W^H$.}
		\State{Perform a rank test for $\Sigma$, i.e., find $0 < k \le l$ such that 
		$$\sigma_1 \ge \sigma_2 \ge \ldots tol_{rank} > \sigma_{k+1} \approx \ldots \approx \sigma_l \approx 0.$$ }
		\If{$k = l$} {$l = l + 1$ and go to line 1;}
		\Else{ let $V_0 = V(1:N, 1:k), W_0 = W(1:l, 1:k)$ and $\Sigma_0 = \diag(\sigma_1, \ldots, \sigma_k)$.}
		\EndIf
		\State{Compute $B = V_0^H A_1 W_0 \Sigma_0^{-1}$.}
		\State{Solve the eigenvalue problem for $B$: $B S = S \Lambda$, where $S=(v_1\vert v_2\vert \ldots \vert v_k)$.}
		\State{For all $j$ with $||T(\lambda_j) v_j || \le tol_{res}$ and $\lambda_j \in \inter(\Gamma)$ accept $\lambda_j$ as nonlinear eigenvalue and $v_j = V_0 s_j$ as eigenvector.}
		\end{algorithmic}
	\label{algo2}
\end{algorithm}

A few further specifications and comments:
\begin{enumerate}
\item
In line 2, we choose $\hat{V} = (I_l\vert O)^T$ with $I_l$ the $l\times l$ identity matrix and $O$ an $l \times (N-l)$ matrix of zeros.
\item
The integrands in line 3 are found by solution of linear systems $T(z) x = \hat{V}$ for $x$.
\item
For the linear systems, SVDs and linear eigenvalue problems we use the Matlab default functions.
\item
We choose as contours circles with the center $\mu$ and radius $R$. Then, for $\varphi(t) = \mu + R \exp\left(\frac{2 \pi i j}{L}\right)$,
$A_0$ and $A_1$ can be approximated by quadrature with $L$ nodes,
		\begin{eqnarray*}
			 A_{0, L} &=& \frac{R}{L} \sum_{j = 0}^{L-1} T(\varphi(t_j))^{-1} \hat{V}  \exp\left(\frac{2 \pi i j}{L}\right), \\
			 A_{1, L} &=& \mu A_{0, L} + \frac{R^2}{L} \sum_{j = 0}^{L-1} T(\varphi(t_j))^{-1} \hat{V} \exp\left(\frac{4 \pi i j}{L}\right).		
		\end{eqnarray*}
\item
As $\mu$ can be large, for computational stability it is better to shift the coordinates to $\tilde{z} = z - \mu$.
\end{enumerate}

\section{Numerical tests}
\label{numerical_solution}

In this section, we discuss the application of the methods in Sections \ref{sec:sepvar} and \ref{NEP} for solving (\ref{Jc_eq_0}) and give numerical tests.

\subsection{Matrix assembly}
\label{subsec:assem}

	To compute the elements of $T$, we have to evaluate the integral in (\ref{J_mn}). We notice that 
	\begin{multline*}
	f_{mn}(\varphi') = \sin{(k_m \varphi')} \sin{(k_n \varphi')} e^{(k_m + k_n) Z(\varphi')} {}_{2}F_1 \left(\lambda, 1 - \lambda, 1 + k_m, \frac{e^{2 Z(\varphi')}}{1+e^{2 Z(\varphi')}}\right) \times \\ \times {}_{2}F_1 \left(\lambda, 1 - \lambda, 1 + k_n, \frac{e^{2 Z(\varphi')}}{1+e^{2 Z(\varphi')}} \right)
	\end{multline*}
	is analytic on $\{|z| < 1\}$ and $f_{mn}'(0) = f_{mn}'(\bar{\omega})$, where $\bar{\omega} = \arccos(-\rho_{xy})$. Then, according to \cite{Trefethen_trapz}, the trapezoidal quadrature rule with $K$ nodes converges with order $O(K^{-4})$ with small constant.
	
	An advantage of using the standard trapezoidal rule compared to other, adaptive quadrature rules is that on a fixed grid we can precompute the functions
	\begin{equation*}
		f_{n}(\varphi') = \sin{(k_n \varphi')} e^{k_n Z(\varphi')}  {}_{2}F_1 \left(\lambda, 1 - \lambda, 1 + k_n, \frac{e^{2 Z(\varphi')}}{1+e^{2 Z(\varphi')}} \right).
	\end{equation*}
Therefore, we use only $N$ instead of $N^2$ computations of hypergeometric functions, which is an expensive operation.

The elements of $T(\lambda)$ grow fast in $n$ and $m$ due to the term $\exp\left((k_n + k_m) Z(\varphi)\right)$, and for large $N$ can become highly inaccurate.
We will use the following straightforward result to improve the condition of $T$.

\begin{proposition}
	\label{lemma_exp_transform}
	Consider non-singular square matrices $V$ and $W$. Define $\tilde{T}(\lambda) = V T(\lambda) W$. Then, $\lambda$ is a nonlinear eigenvalue of $T(\lambda)$ if and only if it is a nonlinear eigenvalue of $\tilde{T}(\lambda)$.
\end{proposition}
Define the matrices as $V = W = D$ with
\begin{equation*}
	D = \diag(e^{-\alpha k_1}, e^{-\alpha k_2}, \ldots, e^{-\alpha k_N}).
\end{equation*}
Then, the elements of $\tilde{T}(\lambda) = D T(\lambda) D$ are
\begin{multline}
	\label{exp_transform}
	\tilde{T}(\lambda)_{m n} = \int \limits_0^{\bar{\omega}} \sin{(k_m \varphi')} \sin{(k_n \varphi')} e^{(k_m + k_n) (Z(\varphi') - \alpha)} {}_{2}F_1 \left(\lambda, 1 - \lambda, 1 + k_m, \frac{e^{2 Z(\varphi')}}{1+e^{2 Z(\varphi')}}\right) \\ \times {}_{2}F_1 \left(\lambda, 1 - \lambda, 1 + k_n, \frac{e^{2 Z(\varphi')}}{1+e^{2 Z(\varphi')}} \right) \, d \varphi'.
\end{multline}

 Transformation (\ref{exp_transform}) helps to decrease the elements proportionally to the growth rate, where $\alpha$ is chosen in the tests as maximum or average value of $Z(\varphi)$.

%

The standard Matlab hypergeometric function is very slow. Therefore, we implemented our own function in C and inserted it into Matlab as mex function.

%

\subsection{Computation of eigenvalues}

Now we consider the application of the numerical methods from Section \ref{NEP} to the truncated problem (\ref{Jc_eq_0}).

\paragraph{Chebyshev approximation of smallest eigenvalue of $T(\lambda)$ (Section \ref{subsec:cheby1}).}
For this method the application is straightforward, using Cholesky factorisation and {\textit{chebfun}} (\cite{Trefethen}) to find the local minima.

\paragraph{Chebyshev polynomial interpolation (Section \ref{subsec:cheby2}).}
Here, we consider 
complex eigenvalues
and Chebyshev interpolation of $T(\lambda)$ is implemented using {\textit{chebfun}} (\cite{Trefethen}). The resulting generalised  linear eigenvalue problem is solved by the default method in Matlab.

\paragraph{Contour integral method (Section \ref{subsec:contour}).}
To find all (complex) eigenvalues, we have to pick suitable centres and radii to apply Algorithm \ref{algo2}.
Based on the empirical observation that the distribution of the eigenvalues is roughly uniform (see Table \ref{eigenval_table} below), 
we can choose a fixed radius and increase the centre by slightly under twice that amount to find the next eigenvalue. We discard any eigenvalues found previously due to the overlap of
contours.  This is summarised in Algorithm \ref{algo3}.
\begin{algorithm}
	\caption{Contour integral algorithm for nonlinear eigenvalue problem}
	\begin{algorithmic}[1]
		\State{Choose $\mu = \mu_0$ and $R$ as discussed above.}
		\While{$\mu \le \mu_{max}$}
			\State{Use Algorithm \ref{algo2} with $\Gamma = C(\mu, R)$, a circle with centre $\mu$ and radius $R$.}
			\State{ $\mu = \mu + 2R$ }
		\EndWhile
	\end{algorithmic}
	\label{algo3}
\end{algorithm}	

	In Tables \ref{eigenval_table} and \ref{eigenval_time_table},
	we compare the approximate nonlinear eigenvalues and the computational time  for different methods (from Section \ref{NEP}) 
	in the semi-analytical approach 
	with those from the finite element method in \cite{LiptonSav}.
	Consider parameters $\rho_{xy} = 0.8, \rho_{xz} = 0.2, \rho_{yz} = 0.5$. For the semi-analytical method we take $N= 20$ for the  approximation of the infinite sum in (\ref{Jc_eq_inf}), $K=1000$ and $L=500$ quadrature nodes.  
	
\begin{table}[H]
	\begin{center}
		\begin{tabular}{| c | c | c | c | c |}
			\hline
			EV & Finite elements & Chebyshev approximation & Contour integrals / \\
			& &   of smallest eigenvalue of $T(\lambda)$ & Chebyshev interpolation \\
			\hline
			$\Lambda_1^2$ & 5.232 & 5.228 &  5.229 (0.028) \\
			$\Lambda_2^2$ & 11.805 & 11.787 & 11.787 (0.076) \\
			$\Lambda_3^2$ & 16.313 &16.285 &  16.284 (0.068) \\
			$\Lambda_4^2$ & 21.204 & 21.149 &  21.147 (0.151) \\
			$\Lambda_5^2$ & 26.184 & 26.112 & 26.109 (0.188) \\
			$\Lambda_6^2$ & 33.392 &  33.230 & 33.261 (0.215) \\
			$\Lambda_{15}^2$ & 72.911 & 72.841 &  72.708 (0.731) \\
			$\Lambda_{30}^2$ & 138.087 & 139.421 &  139.391 (1.331) \\
			\hline
		\end{tabular}
		\caption{Eigenvalues for parameters $\rho_{xy} = 0.8, \rho_{xz} = 0.2, \rho_{yz} = 0.5$. In the last column, the imaginary part is in brackets.As, the contour integral method and Chebyshev polynomial interpolation gave identical results up to 5 digits, 
		we unite them in one column.}
		\label{eigenval_table}
	\end{center}
\end{table}
\begin{table}[H]
	\begin{center}
		\begin{tabular}{|  c | c | c |}
			\hline
			  Chebyshev approximation  & Chebyshev interpolation of $T(\lambda)$ & Contour integrals\\
			  of smallest eigenvalue of $T(\lambda)$ & & \\
			\hline
			561.17 s & 60.27 s & 220.79 s \\
			\hline
		\end{tabular}
		\caption{Computation time of first 30 eigenvalues.}
		\label{eigenval_time_table}
	\end{center}
\end{table}
{\color{black}
We also compare the numerically computed nonlinear eigenvalues to the analytic expressions in the case $\rho_{xy} = \rho_{xz} = \rho_{yz} = 0$ (see Appendix \ref{app:zerocorr}),  in Table \ref{eigenval_table_zero_corr}. We can see that the semi-analytical method is vastly superior to the finite element method. The semi-analytical method has the advantage in this case that there exists a finite basis and the truncation for the nonlinear eigenvalue problem after sufficiently many, say $N$ terms does not change the solution (see Appendix \ref{app:zerocorr}). Here, $N$ is chosen large enough so that the remaining errors come
predominantly from the approximate solution of the linear eigenvalue problem in the case of the Chebyshev method and the singular value decomposition in the contour integral
method.
}

\begin{table}[H]
	\begin{center}
	{\color{black}
		\begin{tabular}{| c | c | c | c | c | c | }
			\hline
			EV & Value & Finite elements & Chebyshev approx. & Chebyshev  &  Contour integrals \\
			& &   & of sm. eig of $T(\lambda)$ & interpolation& \\
			\hline
			$\Lambda_1^2$ & 12.0 & 0.018 & $4.1 \times 10^{-15} $&  $3.6 \times 10^{-12} $ &  $1.3 \times 10^{-12}$ \\
			$\Lambda_2^2$ & 30.0 &  0.097 & $9.5 \times 10^{-14}$ &  $1.1 \times 10^{-11}$ & $3.4 \times 10^{-12}$\\
			$\Lambda_3^2$ & 30.0 &0.105 & $9.5 \times 10^{-14}$ &  $1.1 \times 10^{-11}$ & $3.4 \times 10^{-12}$\\
			$\Lambda_4^2$ & 56.0 & 0.309 & $7.7 \times 10^{-12}$ &  $6.0 \times 10^{-11}$ & $6.1 \times 10^{-11}$\\
			$\Lambda_5^2$ & 56.0 & 0.353 & $7.7 \times 10^{-12}$ & $6.0 \times 10^{-11}$ & $6.1 \times 10^{-11}$\\
			$\Lambda_7^2$ & 90.0 & 0.767 &  $1.0 \times 10^{-10}$ & $5.4 \times 10^{-10}$ & $7.8 \times 10^{-10}$\\
			$\Lambda_{15}^2$ & 132.0 & 2.311 &$2.3 \times 10^{-5}$ &  $1.8 \times 10^{-9}$ & $4.5 \times 10^{-8}$\\
			$\Lambda_{30}^2$ & 306.0 & 8.896 & $1.3 \times 10^{-2}$ &  $4.1 \times 10^{-8}$ & $1.0 \times 10^{-6}$ \\
			\hline
		\end{tabular}
		
		\caption{
		{\color{black}
		Error of eigenvalues for $\rho_{xy} = 0.0, \rho_{xz} = 0.0, \rho_{yz} = 0.0$. We present the difference between eigenvalues computed by different methods and the analytical solution.}}
		\label{eigenval_table_zero_corr}
		}
	\end{center}
\end{table}

	From the numerical experiments we can conclude that the method based on the smallest linear eigenvalue is slow. The methods based on contour integrals and Chebyshev interpolation show the best results. From a computational point of view, Chebyshev interpolation is faster than the contour integral method, because for the latter we have to run the computations several times on different circles, as well as compute numerical integrals along contours.

The complexity of the Chebyshev interpolation method consists of two steps. The first step is the interpolation of $T(\lambda)$ in $\lambda$ using Chebyshev polynomials as in \eqref{pol_eigval} to obtain
a polynomial eigenvalue problem. This can be performed in $O(N^2 m)$, where $N$ is the number of terms in the eigenvalue expansion and $m$ is the dimension of the Chebyshev basis. The second step is to solve the polynomial eigenvalue problem, for which the complexity equals that of the linear eigenvalue decomposition of a matrix of size $Nm \times Nm$. Thus, the complexity of the second step is proportional to that of matrix multiplication, i.e., $O(N^3 m^3)$ in practice.\footnote{Theoretically, algorithms with $O(N^{\omega} m^{\omega})$ for $2 < \omega < 2.37$ are available (see \cite{FastLinAlg}), but the constant is found to be too large to be competitive for the matrix size relevant here.}

We give an empirical analysis of the CPU time in Table \ref{empirical_table}. Since the first step has a higher constant factor, it dominates for small $m$, while for large $m$ the second step dominates.
%
%

\begin{figure}
	\begin{center}
				\subfloat[Eigenvector 1: $\Lambda_1^2 = 5.2$]{\includegraphics[width=0.5\textwidth]{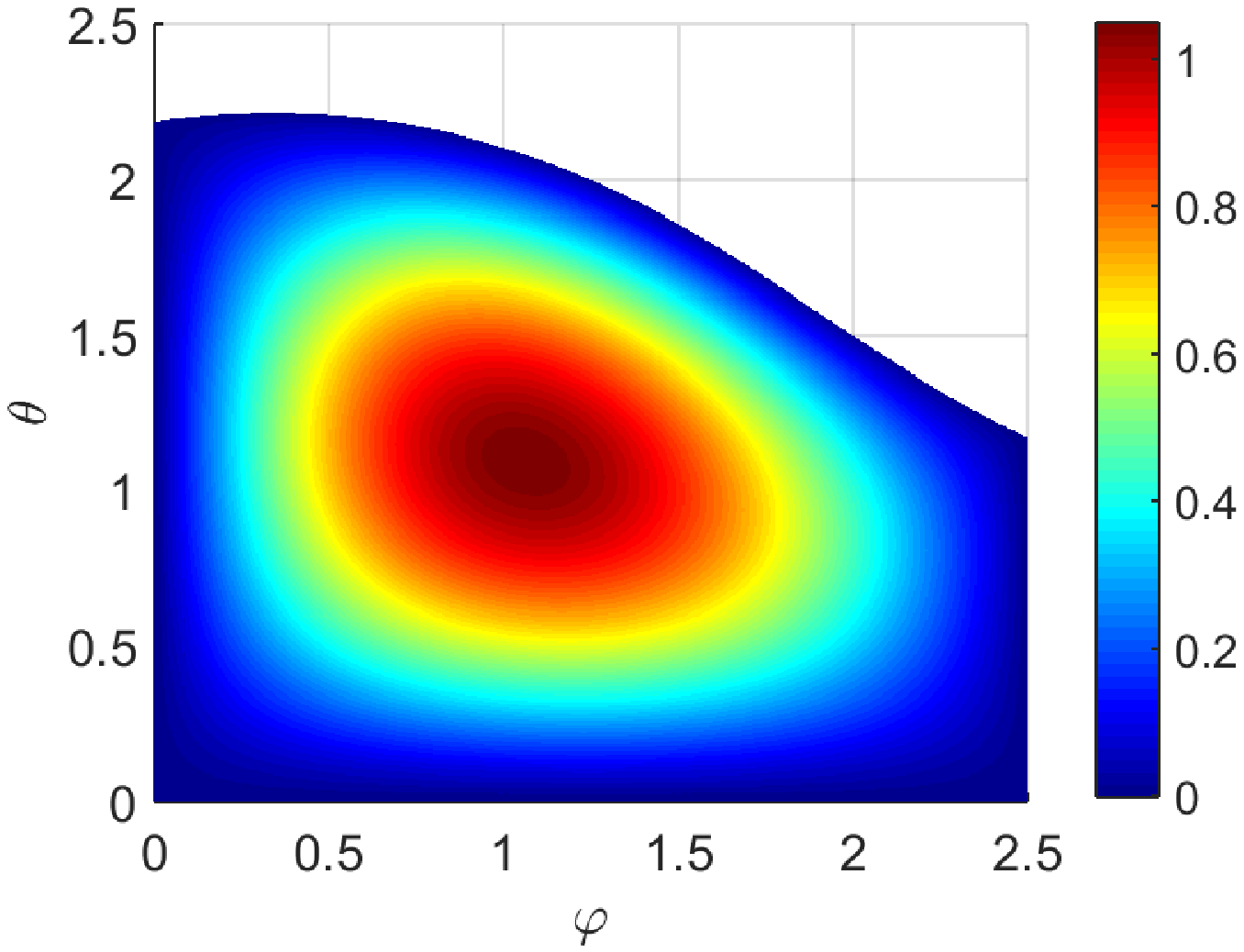}}
				\subfloat[Eigenvector 2: $\Lambda_2^2 = 11.8$]{\includegraphics[width=0.5\textwidth]{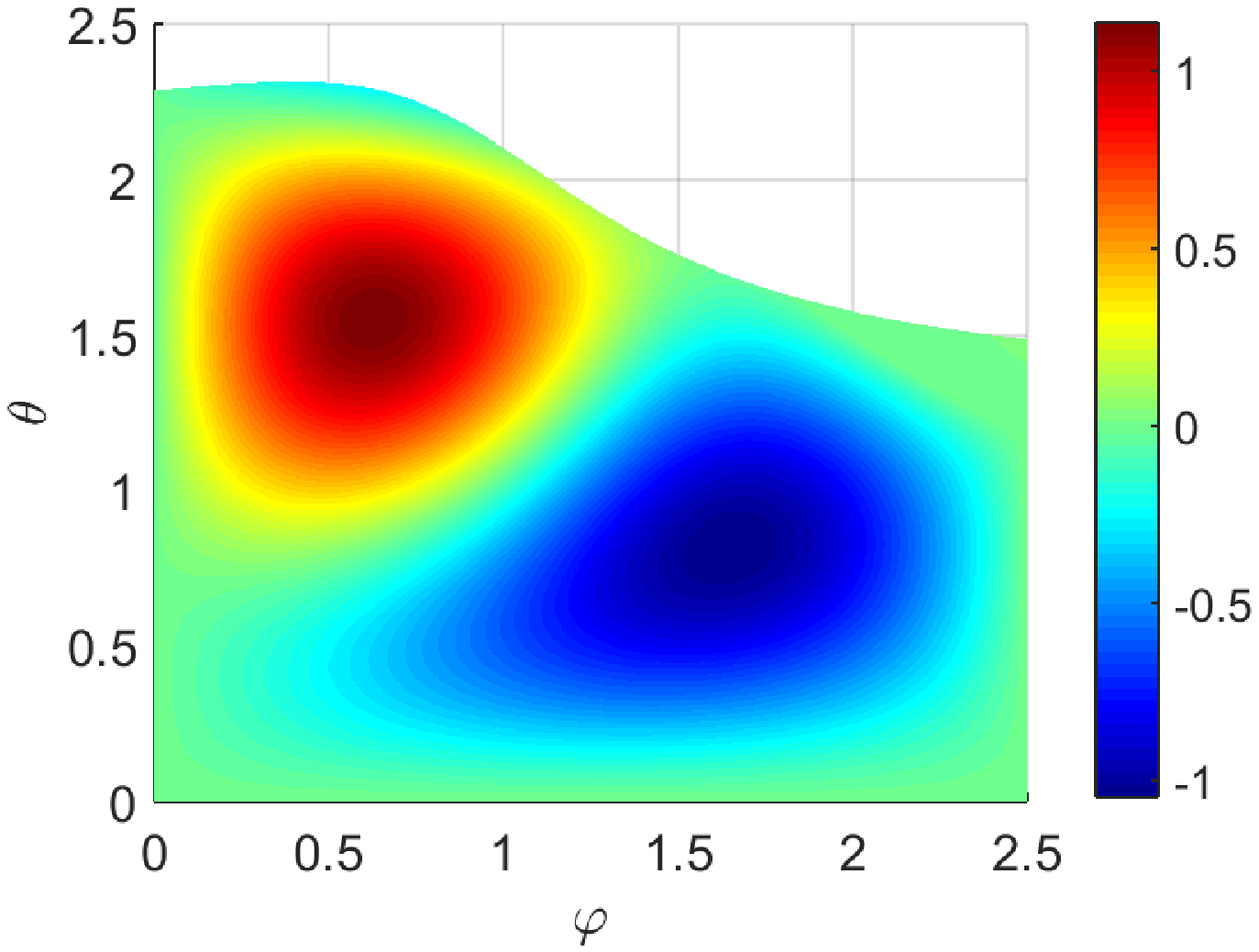}}\\
				\subfloat[Eigenvector 3: $\Lambda_3^2 = 16.3$]{\includegraphics[width=0.5\textwidth]{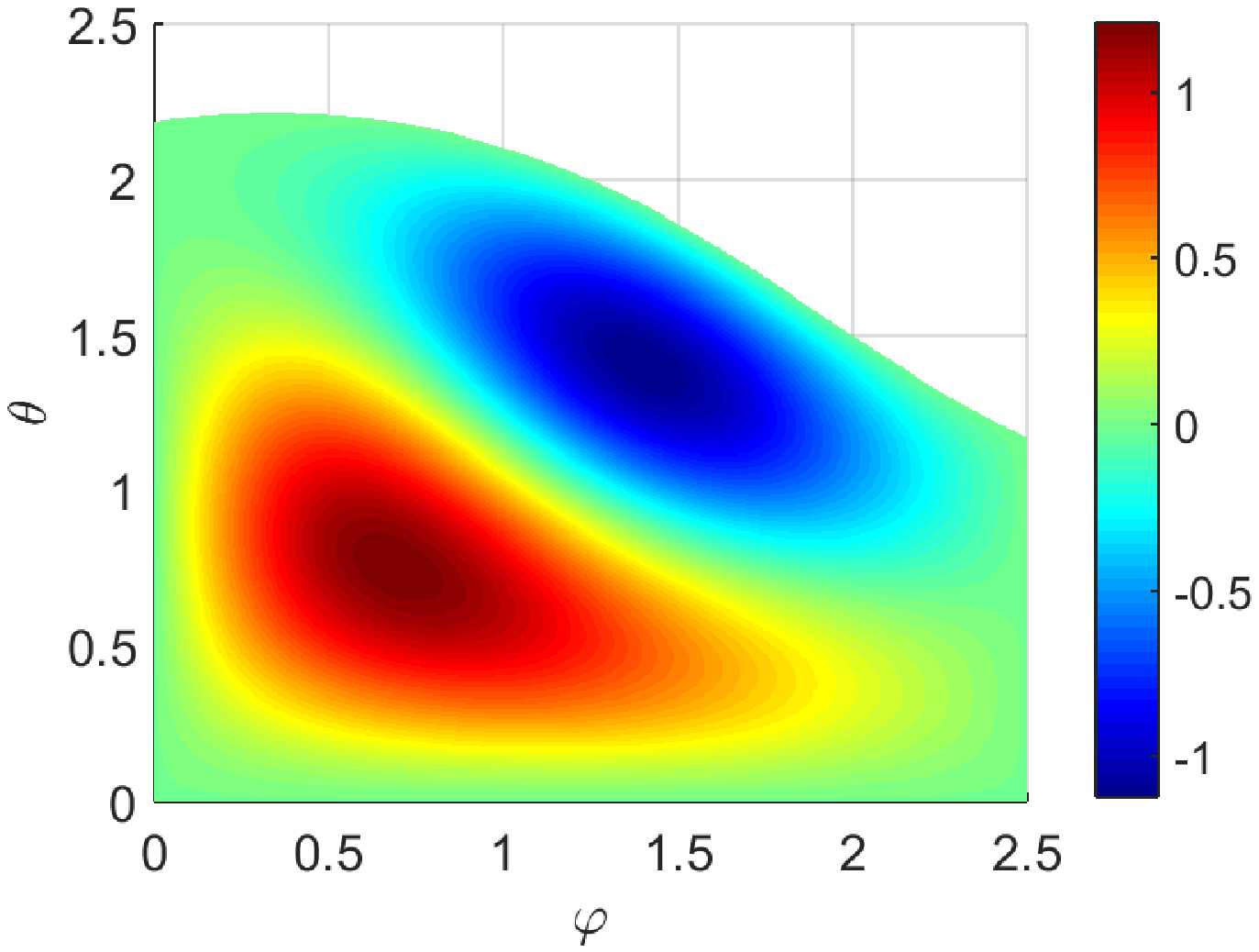}}
				\subfloat[Eigenvector 4: $\Lambda_4^2 = 21.1$]{\includegraphics[width=0.5\textwidth]{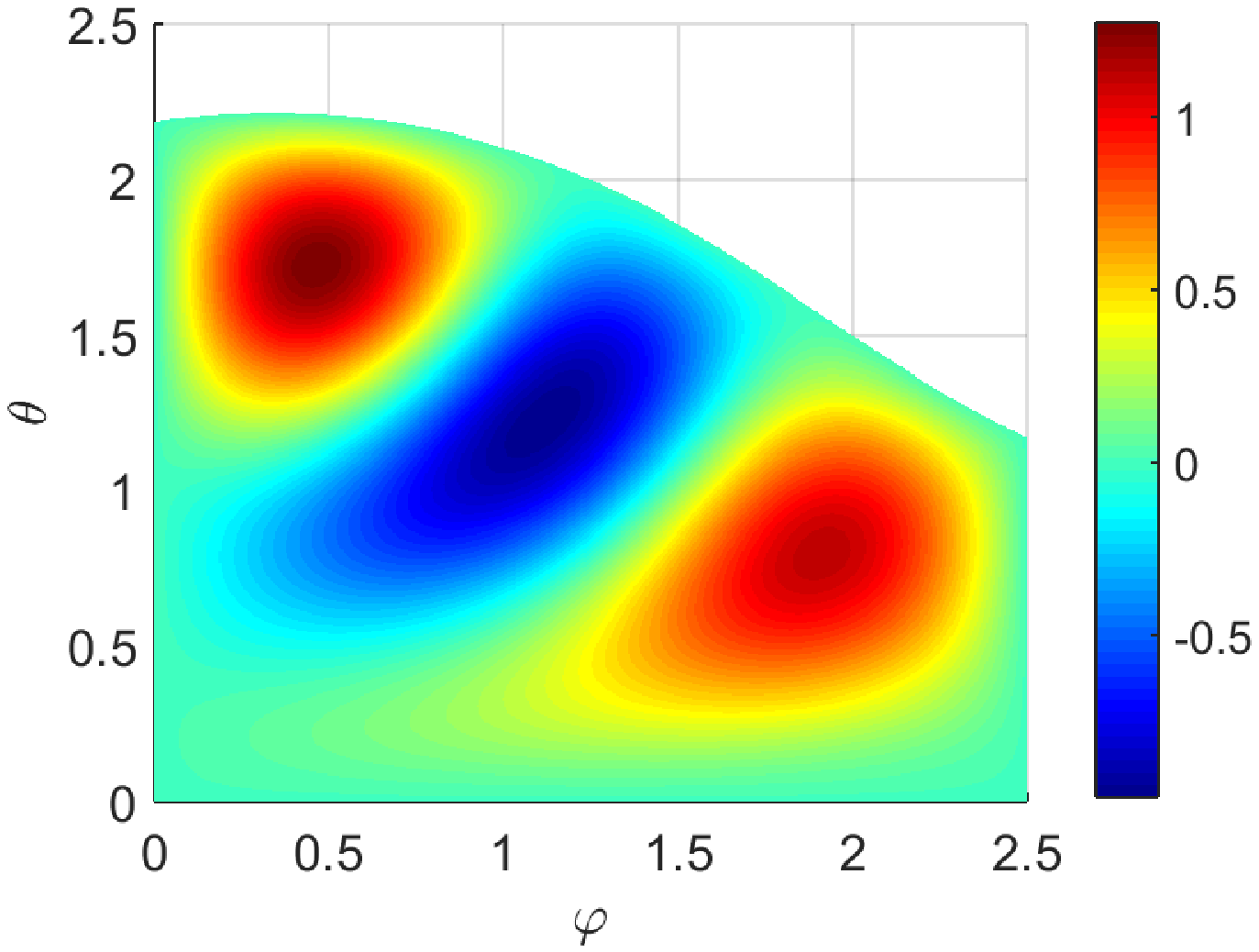}}\\
				\subfloat[Eigenvector 8: $\Lambda_8^2 = 39.3$]{\includegraphics[width=0.5\textwidth]{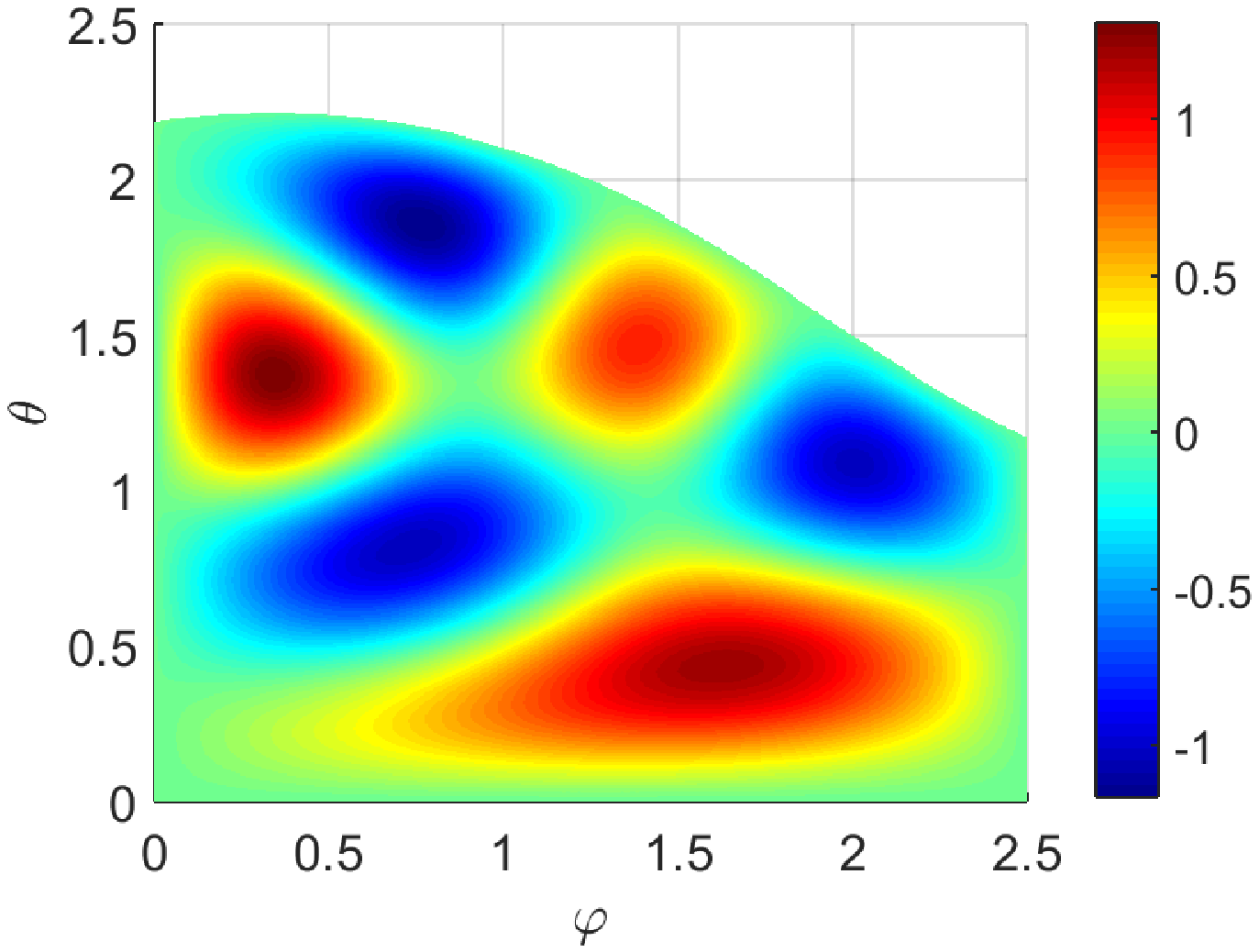}}
				\subfloat[Eigenvector 30: $\Lambda_{30}^2 = 139.4$]{\includegraphics[width=0.5\textwidth]{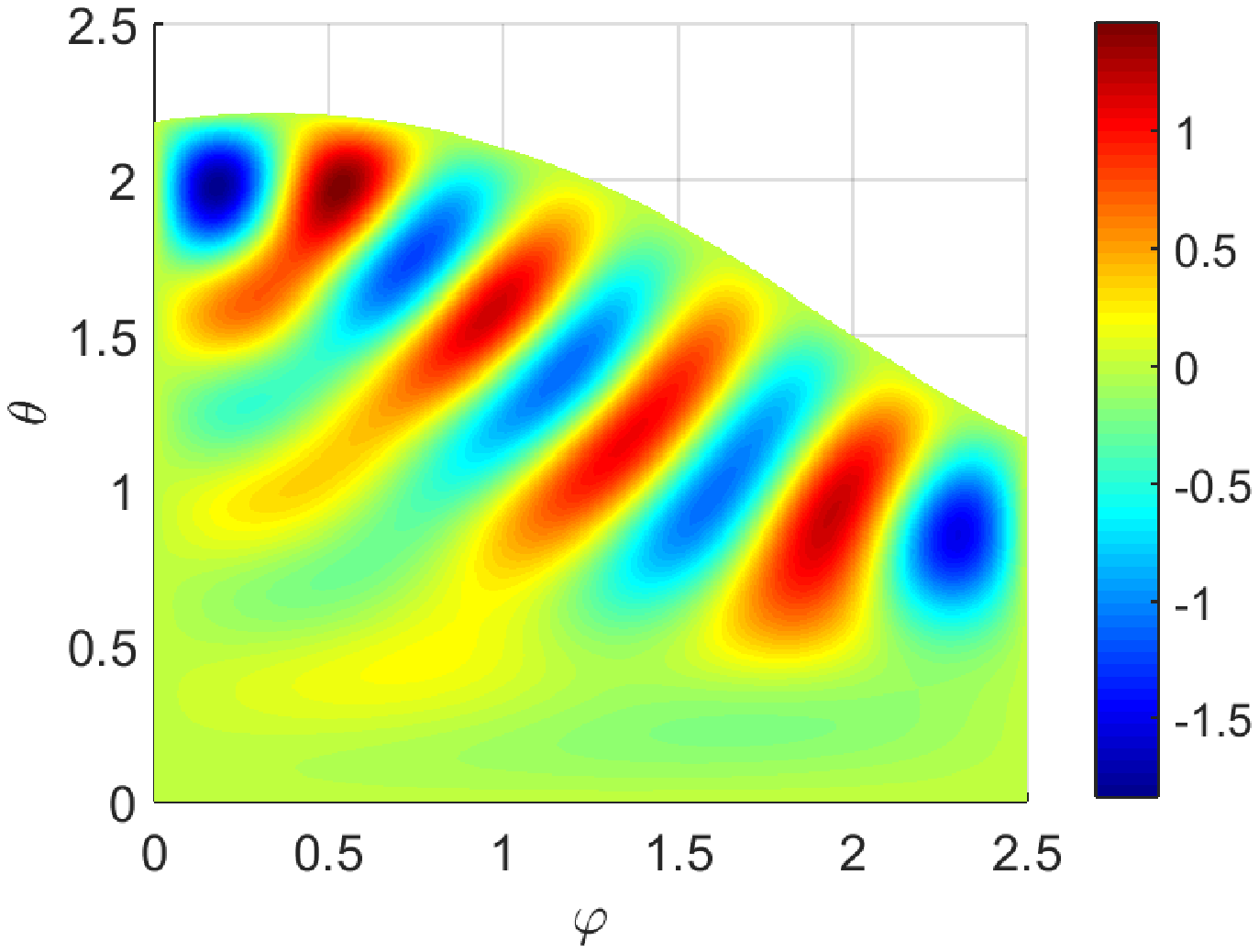}}\\
	\end{center}		
	\vspace{-20pt}
	\caption{An example of eigenvectors and corresponding eigenvalues for the domain obtained for $\rho_{xy} = 0.8, \rho_{xz} = 0.2, \rho_{yz} = 0.5$.}
 	\label{eigenvec_example}
\end{figure}

Once we have computed the eigenvalues and the corresponding eigenvectors, we can compute $\Psi(\varphi, \zeta)$ using expression (\ref{Psi_expression}). An example of eigenvectors is given in Figure \ref{eigenvec_example}, 
shown in the $(\varphi, \theta)$-plane.

\subsection{Convergence of expansion and computational speed}

We now focus on the most efficient method, the Chebyshev interpolation method from Section \ref{subsec:cheby2}.
To explore the convergence numerically, we compare our results for the Green's function with the analytical solution
available for the special correlation case $\rho_{xy} = -\cos{\left(\frac{\pi}{3}\right)}, \rho_{xz} = -\cos{\left(\frac{\pi}{3}\right)}, \rho_{yz} = 0$
using a method of images (\cite{escobar2013three}).

{\color{black}
 It is worth pointing out that we should not expect uniform convergence in $(t', x', y', z')$, especially as $t'\downarrow t$, because the initial condition is a Dirac delta.
 In applications, the quantity of interest is usually an integral in the space variables of the Green's function for $t' > 0$. It is well-known that for $t' > 0$ the Green's function is smoothed. In our experiments, we consider the spatial $L_2$ norm (for fixed $t'$) for comparison.
}

Our truncated Green's function approximation has the form
\begin{multline*}
	\label{greens_approx}
		G(\tau, r', \varphi', \theta', r, \varphi, \theta)  =  \frac{e^{-\frac{r'^2+r^2}{2 \tau}}}{\tau \sqrt{r' r}}\sum_{l = 1}^{M} I_{\lambda_l} \left(\frac{r' r}{\tau}\right) \times \\
		\times \left(\sum_{n = 1}^{N} c_n^l \sin{\left(k_n \varphi' \right)}  \tan^{k_n}\left(\frac{\theta'}{2}\right) {}_{2}F_1 \left(\lambda_l, 1 - \lambda_l, 1 + k_n, \sin^2 \frac{\theta'}{2} \right)\right) \times \\
		\times  \left(\sum_{n = 1}^{N} c_n^l \sin{\left(k_n \varphi \right)} \tan^{k_n}\left(\frac{\theta}{2}\right) {}_{2}F_1 \left(\lambda_l, 1 - \lambda_l, 1 + k_n, \sin^2 \frac{\theta}{2} \right)\right). 
\end{multline*}

{\color{black}
We have several numerical parameters:
\begin{enumerate}[(i)]
\item
$M$, the number of terms in the superposition solution of (\ref{3d_eq_spherical}), i.e., the number of nonlinear eigenvalues and eigenvectors in 
(\ref{Jc_eq_inf}) we approximate;
\item
$N$, the dimension of the nonlinear eigenvalue problem (\ref{Jc_eq_0}) approximating the infinite-dimensional problem (\ref{Jc_eq_inf});
\item
$m$, the dimension of the Chebyshev basis in (\ref{pol_eigval}) to approximate the nonlinear eigenvalue problem by a polynomial one;
\item
$K$, the number of quadrature points to approximate the integral in (\ref{J_mn}); see also Section \ref{subsec:assem}.
\end{enumerate}

We expect convergence of the approximation as those four parameters go to infinity.
This involves in particular that $c_l^n$, combined with the terms they are multiplied by, go to zero sufficiently fast as $n,l\rightarrow 0$. Moreover, one needs that for fixed $n,l$ the $c_l^n$ and $\lambda_l$, which implicitly depend on $N$ through the truncation of the infinite-dimensional eigenvalue problem and on $m$ and $K$ by the approximations made 
to the finite-dimensional eigenvalue problems, converge as $N,m,K \rightarrow \infty$.
}

In absence of theoretical error estimates, we test this empirically.
In our experiments, we choose $K = 1000$ nodes for the trapezoidal rule for \eqref{J_mn};  very good tolerance is already achieved with much fewer nodes, however, even for $K = 1000$ the computational time for these integrals is not very significant (about 10-15\% of the total eigenvalue computation procedure).

We now provide three experiments: in each, we fix two of the remaining parameters ($M$, $N$, $m$) to be large enough and vary the third parameter to compare the Green's function with the analytical solution.
For $N = 30, M = 50, m = 100$, in our experiments, further increase of either parameter does not affect the Green's function's value beyond the accuracy required here.

The results and computational times are given in Table \ref{empirical_table} below\footnote{{\color{black} \label{footnote_label} In order to approximate the $L_2$ error, we compute the Green's function on a grid
$\{(x'_i, y'_j, z'_k) = (i \Delta x, j \Delta y, k \Delta z)\}$ covering some $[0, x^{max}] \times [0, y^{\max}] \times [0, z^{max}]$ and then approximate the error by numerical quadrature by
$\errL^2 \approx {\sum_{i, j, k} \left(G(x'_i, y'_j, z'_k, x, y, z) - G^{*}(x'_i, y'_j, z'_k, x, y, z) \right)^2 \Delta x \Delta y \Delta z} $, where $G(\cdot)$ is the semi-analytical or numerical solution and $G^{*}(\cdot)$ the analytical solution from \cite{escobar2013three}. }}.
Recall that the infinite-dimensional nonlinear eigenvalue problem (\ref{Jc_eq_inf}) is ultimately approximated by the $(m N)$-dimensional generalized linear eigenvalue problem
for (\ref{pencil}). To investigate the convergence in the number of terms in the expansion, $M$, we choose the $M$ smallest out of the resulting $m N$ eigenvalues and corresponding eigenvectors. As the main computational cost comes from the solution of the eigenvalue problem, which is here identical for all $M$, we do not report computational times for different $M$.

{\color{black} In comparison, using the finite element method from \cite{LiptonSav} with a 1500 point mesh, we get the precision $3.1 \times 10^{-5}$ with CPU time 621s.
This accuracy is comparable with that of the semi-analytical solution for $m$ around 15 and  $N$ around 7, which is computable within a fraction of a second.
Note that the approach of \cite{LiptonSav} requires the solution of a 1500-dimensional (determined by the number of finite elements) generalized linear eigenvalue problem. In our
tests, however, the cost was dominated by the construction of the locally refined finite element mesh.}

\begin{table}[H]
	\begin{center}
		\begin{tabular}{|  c | c | c | c | c | c |}
			\hline \hline
			$m$ & 10& 20 & 50 & 80 & 100 \\
			\hline
			Error & $2.52 \times 10^{-4}$ & $7.73 \times 10^{-6}$ & $6.21 \times 10^{-7}$ & $1.76 \times 10^{-7}$ & $8.60 \times 10^{-8}$   \\
			CPU time (s) & $79.9$ & $94.8$ & $256.5$ & $604.1$ & $1070.1 $  \\
			\hline \hline
			$N$ & 5& 10 & 20 & 25 & 30 \\
			\hline
			Error & $7.92 \times 10^{-5}$ & $3.46 \times 10^{-6} $ & $2.68 \times 10^{-7}$  & $8.60 \times 10^{-8} $ & $8.60 \times 10^{-8}$  \\			
			CPU time (s) & $9.9$ & $56.6$ & $333.1$ & $667.4$ & $1070.1$  \\
			\hline \hline
			$M$ & 5 & 10 & 20 & 40 & 50 \\
			\hline
			 Error & $4.87 \times 10^{-4}$& $1.19 \times 10^{-5}$ & $9.71 \times 10^{-7}$ & $1.02 \times 10^{-7}$  & $8.60 \times 10^{-8}$   \\
			 \hline \hline
		\end{tabular}
		\caption{$L_2$ norm  of the Green's function's error for different $m$ ($N = 30, M = 50$), $N$ ($m = 100, M = 50$), and $M$ ($m = 100, N = 30$).}
		\label{empirical_table}
	\end{center}
\end{table}
%

As a further benchmark, we present the results for a second order ADI finite-difference method (the Hundsdorfer-Verwer scheme; see \cite{HV}) in Table \ref{FD_x}
below\footnotemark[\value{footnote}].
We use up to $N_X=100$ points in each spatial direction and $N_T=200$ points in time.

{\color{black}
The numerical initial condition is chosen as
\begin{equation*}
	G(0, x_1^{i}, x_2^{j}, x_3^{k}, x_1, x_2, x_3) =  \left\{
	\begin{array}{rl}
		\frac{1}{\Delta x_1 \Delta x_2 \Delta x_3}, & \quad x_1^{i} = x_1, x_2^{j} = x_2, x_3^{k} = x_3, \\
		0, & \quad \text{otherwise}.
	\end{array}
	\right.
\end{equation*}
in order to approximate the Dirac delta, with the mesh constructed in such a way that $(x_1, x_2, x_3)$ coincides with a mesh point.
This is a special case of the approximations to non-smooth initial conditions analysed numerically in \cite{pooley2003} and by means of Fourier methods in \cite{reisinger2014impact}.
}

The data in Table \ref{FD_x} show second order convergence in both $N_X$ and $N_T$.

\begin{table}[H]
	\begin{center}
		\begin{tabular}{|  c | c | c | c |}
			\hline \hline
			$N_X$ & 25 & 50 & 100\\
			\hline
			Error & $4.44 \times 10^{-4}$ & $1.04 \times 10^{-4} $ & $2.65 \times 10^{-5} $  \\
			CPU time (s) & $30.8$ & $274.3$ & $2239.1$\\
			\hline \hline
			$N_T$ & 50& 100 & 200  \\
			\hline
			Error & $3.41 \times 10^{-4}$ & $9.23 \times 10^{-5} $ & $2.65 \times 10^{-5} $  \\
			CPU time (s) & $723.2$ & $1156.6$ & $2239.1$  \\
			\hline \hline			
		\end{tabular}
		\caption{$L_2$ norm of the Green's function's error for the ADI method with different  $N_X$ ($N_T=200$) and $N_T$ ($N_X=100$).}
		\label{FD_x}
	\end{center}
\end{table}

This test demonstrates the superior accuracy of the semi-analytical solution for relatively few terms in comparison to the ones
obtained for the ADI scheme  on the highest computationally feasible levels. 

It is worth pointing out though that the semi-analytical expression gives the density in a single point, while the finite difference solution is simultaneously obtained on a whole mesh.
This is useful, for example, to compute survival probabilities, where one needs to integrate the density (see next section).


\section{Applications to default model with mutual liabilities}

Diffusions with absorption play a central role in structural credit models.
In particular,
we investigate in this section such a model of an interconnected banking network with mutual liabilities as in \cite{Lipton2015}.
Several papers have considered specific cases of this model. For example, \cite{LiptonItkin2015} considered the special case of two banks without jumps, \cite{Kaushansky2017} considered the case of two banks with negative exponential jumps, and \cite{LiptonItkin2014} considered the cases of two and three banks with a fairly general correlated L\'evy process in the jump component.

Here, we study the case of three banks without jumps. The case of three counterparties is of practical importance as it allows us to compute bilateral Credit and Debt Value Adjustments. In the following, we briefly introduce the framework and show how to compute the joint survival probability, as well as CVA and DVA for a CDS.
\subsection{Extended default model with mutual liabilities for three banks}
We assume that assets are driven by geometric Brownian motion with drift
\begin{equation*}
	\frac{d A_{i,t}}{A_{i,t}} = \mu_i \, d t + \sigma_i \, d W^i_t, \quad 1\le i \le  3, 
\end{equation*}
where $\mu_i$ is the growth rate, $\sigma_i$ is the corresponding volatility, $W^i$ are correlated standard Brownian motions
with correlation $(\rho_{ij})$.
The liabilities are assumed to be deterministic with the same growth rate,
\begin{equation*}
	\frac{d L_i}{L_i} = \mu_i \, d t, \quad \frac{d L_{ij}}{L_{ij}} = \mu_i \, d t, \quad 1\le i, j \le  3,
\end{equation*}
where $L_i$ are the external liabilities of the $i$-th bank, and $L_{ij}$ are interbank liabilities between bank $i$ and bank $j$.

We introduce the default boundaries as in \cite{BlackCox}
\begin{equation*}
	\Xi_i = \left\{
	\begin{aligned}
	& R_i ( L_i + \hat{L}_i ) - \hat{A}_i \equiv \Xi_i^{<},  \quad & t < T, \\
	& L_i + \hat{L}_i - \hat{A}_i \equiv \Xi_i^{=},  & t = T,
	\end{aligned}
	\right.
\end{equation*}
where $0 \le R_i \le 1$ is the recovery rate and
\begin{equation*}
	\hat{A}_i = \sum_{j \ne i} L_{ji}, \quad \hat{L}_i = \sum_{j \ne i} L_{ij}.
\end{equation*}

If the $k$-th bank defaults at intermediate time $t'$, the surviving banks suffer losses due to any of the $k$-th bank's unpaid net obligations to them and we have to consider a two-dimensional problem with modified boundary conditions.
We then change the indexation of the surviving banks by applying the function
\begin{equation*}
	i \rightarrow i' = \phi^k(i) = 
	\begin{cases}
		i, &i < k, \\
		i - 1, & i > k.
	\end{cases}
\end{equation*}
We also introduce the inverse function
\begin{equation*}
	i \rightarrow i' = \psi^k(i) = 
	\begin{cases}
		i, &i < k, \\
		i + 1, & i \ge k.
	\end{cases}
\end{equation*}

The corresponding default boundaries are modified to
\begin{equation*}
	\Xi_i^{(k)} = 
	\begin{cases}
		 R_{\psi^{(k)}(i)} (L_i^{(k)} + \hat{L}_i^{(k)}) - \hat{A}_i^{(k)},  &t < T, \\
		 L_i^{(k)} + \hat{L}_i^{(k)} - \hat{A}_i^{(k)},  &t = T.
	\end{cases}
\end{equation*}
It is clear that 
\begin{equation*}
	\Delta \Xi_i^{(k)} =  \Xi_i^{(k)} - \Xi_i = 
	\begin{cases}
		 (1 - R_i R_k) \hat{A}_i^{(k)},  &t < T, \\
		 (1 - R_k) \hat{A}_i^{(k)},  &t = T.
	\end{cases}
\end{equation*}
Thus, $\Delta  \Xi^{(k)} > 0$ and the corresponding default boundaries move to the right.

We need to specify the settlement process at time $t = T$. We shall do this in the spirit of \cite{Eisenberg}. Since at time $T$ full settlement is expected, we assume that bank $i$ will pay the fraction $\omega_i$ of its total liabilities to creditors. This implies that if $\omega_i = 1$ the bank pays all liabilities (both external and interbank) and survives. On the other hand, if $0 < \omega_i < 1$, bank $i$ defaults, and pays only a fraction of its liabilities. Thus, we can describe the terminal condition as a system of equations
\begin{equation}
	\label{term_cond}
	\min \left\{A_{i,T} + \sum_{j \ne i} \omega_j L_{ji}, L_i + \sum_{j \ne i} L_{ij}  \right\} = \omega_i \left(L_i + \sum_{j \ne i} L_{ij} \right).
\end{equation}
There is a unique vector $\omega = (\omega_1, \omega_2, \omega_3)^T$ such that the condition (\ref{term_cond}) is satisfied. See \cite{Lipton2015} for details.

For convenience, we introduce normalized dimensionless variables
\begin{equation*}
	\bar{t} = \Sigma^2 t, \quad X^i = \frac{\Sigma}{\sigma_i} \ln \left(\frac{A_i}{\Xi_i^{<}}\right),
\end{equation*}
where
$
	\Sigma = (\sigma_1 \sigma_2 \sigma_3)^{\frac{1}{3}}.
$
Denote also
$\xi_i = -\sigma_i/(2 \Sigma)$ for $1\le i \le  3$.
Applying Itô's formula to $X^i$, we find its dynamics
\begin{equation*}
	d X^i_{\bar{t}} = \xi_i \, d \bar{t} + d W^i_{\bar{t}}.
\end{equation*}
In the following, we omit bars for brevity and
denote the state variable by $(x, y, z) \in \mathbb{R}^3_{+}$. 
The default boundaries change to 
\begin{equation*}
	\mu_i =
	\begin{cases}
		\mu_i^{<} = 0, & t < T, \\
		\mu_i^{=} = \frac{\Sigma}{\sigma_i} \ln \left(\frac{\Xi_i^{=}(t)}{\Xi_i^{<}(t)} \right), & t = T.
	\end{cases}
\end{equation*}

The problems we will consider can all be written in the form
\begin{align}
\label{PDEgen}
		\frac{\partial V}{\partial t} + \mathcal{L} V &= \chi(t, x, y, z), \\
		V(t, 0, y, z) &= \phi_{0, x}(t, y, z), \quad V(t, x, y, z) \underset{x \to +\infty}{\longrightarrow} \phi_{\infty, x}(t, y, z), \\
		V(t, x, 0, z) &= \phi_{0, y}(t, x, z), \quad V(t, x, y, z) \underset{y \to +\infty}{\longrightarrow} \phi_{\infty, y}(t, x, z), \\
		V(t, x, y, 0) &= \phi_{0, z}(t, x, y), \quad V(t, x, y, z) \underset{z \to +\infty}{\longrightarrow} \phi_{\infty, z}(t, x, y), \\
		V(T, x, y, z) &= \psi(x, y, z),
		\label{TCgen}
\end{align}
with suitably defined terminal conditions $\psi$, boundary conditions $\phi$ and right-hand sides $\chi$, and
where the Kolmogorov backward operator is
\begin{equation*}
	\label{kolmogorov_backward}
	\mathcal{L} f = \frac{1}{2}  f_{x x} + \frac{1}{2}  f_{y y} + \frac{1}{2}  f_{z z} + \rho_{12} f_{x y} + \rho_{13} f_{x z} + \rho_{23} f_{y z} +   \xi_1 f_{x} + \xi_2 f_{y} + \xi_3 f_{z}.
\end{equation*}

Once the Green's function has been constructed numerically for a set of model parameters, prices of different derivatives and hedge parameters can be found
as the solution of (\ref{PDEgen})--(\ref{TCgen}) by simple convolution of the Green's function with the appropriate inhomogeneous right-hand sides and boundary data.
It is the strength of the Green's function concept that no further numerical solution of PDEs is required. We illustrate this in the following sections.

\subsection{Joint survival probability}
The corresponding Kolmogorov backward equation for the joint survival probability is
\begin{align*}
		& \frac{\partial Q}{\partial t} + \mathcal{L} Q = 0, \\
		& Q(t, 0, y, z) =  0, \quad Q(t, x, 0, z) = 0, \quad Q(t, x, y, z) = 0, \\
		& Q(T, x, y, z) = \mathbbm{1}_{(x, y, z) \in D_{xyz}},
\end{align*}
where $D_{xyz} = \{ x \ge \mu_x, y \ge \mu_y, z \ge \mu_z \}$ is the subset of $\mathbb{R}^3_{+}$ where all banks survive.

Due to various regulation requirements, assets of large banks cannot drop below their liabilities, which means that their recovery rates $R$ should be close to 1. This implies that the domain $D_{xyz}$ becomes the positive octant $\mathbb{R}^3_{+}$. 

The analytical solution for the survival probability in the positive quadrant (i.e., two dimensions) with non-zero drift was recently found by \cite{LiptonItkin2015}.
Here, we extend this result to three dimensions using the expression for the Green's function from Section \ref{sec:sepvar} and an expansion in Gegenbauer polynomials.

In this section alone, we use $(x,y,z)$ instead of $(x',y',z')$, and $(x_0,y_0,z)$ instead of $(x,y,z)$, to shorten the notation in the integrals,
and similarly for $(r,\varphi, \theta)$.
The solution $Q$ can then be written as
\begin{equation*}
	Q(\tau, x_0, y_0, z_0) = \iiint \limits_{\mathbb{R}_{+}^3} G(\tau, x', y', z', x_0, y_0, z_0) \, dx' dy' dz'.
\end{equation*}
Using \eqref{greens_final} and \eqref{non_drift},
\begin{multline}
	\label{Q_eq}
	Q(\tau, r_0, \varphi_0, \theta_0) = \int \limits_0^{+\infty} \int \limits_0^{\bar{\omega}} \int \limits_0^{\Theta(\varphi)}\exp \left(\xi_{\alpha}(r \sin{\theta} \sin{\varphi} - r_0 \sin{\theta_0} \sin{\varphi_0}) + \right. \\
	\left.  \xi_{\beta}(r \sin{\theta} \cos{\varphi} - r_0 \sin{\theta_0} \cos{\varphi_0}) + \xi_{\gamma}(r \cos{\theta} - r_0 \cos{\theta_0})\right) G(\tau, r, \theta, \varphi) r^2 \sin{\theta} \, d \theta d \varphi d r = \\
	\frac{e^{\kappa}}{\tau \sqrt{r_0}} \sum_{l = 1}^{\infty} \Psi_l(\varphi_0, \theta_0) \int \limits_0^{\bar{\omega}} \int \limits_0^{\Theta(\varphi)} \Psi_l(\varphi, \theta) \sin{\theta} \int_0^{+\infty} e^{- \alpha r^2} r^{\frac{3}{2}} e^{\gamma(\varphi, \theta) r} I_{\lambda_l}(\beta r) \,d r d \theta d \varphi,
\end{multline}
where $\lambda_l = \sqrt{\Lambda_l^2 + \frac{1}{4}}$, $\alpha = \frac{1}{2 \tau}, \beta = \frac{r_0}{\tau}, \gamma(\varphi, \vartheta) = \xi_{\alpha} \sin{\theta} \sin \varphi + \xi_{\beta} \sin{\theta} \cos{\varphi} + \xi_{\gamma} \cos{\theta}$ and $$\kappa = -\xi_{\alpha }r_0 \sin{\theta_0} \sin{\varphi_0} -\xi_{\beta}r_0 \sin{\theta_0} \cos{\varphi_0}  -\xi_{\gamma }r_0 \cos{\theta_0} - \alpha r_0^2.$$

Next, we use the idea of exponent representation via ultra-spherical polynomials from \cite{LiptonItkin2015},
\begin{equation*}
	e^{-Sx} = \Gamma(\nu) \left(\frac{S}{2}\right)^{-\nu} \sum \limits_{k = 0}^{\infty} (-1)^k (\nu + k) I_{\nu + k}(S) C_k^{\nu}(x),
\end{equation*}
where $C_k^{\nu}(x)$ are Gegenbauer polynomials and the parameter $\nu$ can be chosen arbitrarily.

Substituting this representation with $S = \beta R, x = -\gamma(\varphi, \theta)/\beta$ into (\ref{Q_eq}), we get
\begin{multline}
\label{Q_via_two_int}
Q(\tau, r_0, \varphi_0, \theta_0) = 	\frac{e^{\kappa}}{\tau \sqrt{r_0}} \Gamma(\nu) \left(\frac{\beta}{2}\right)^{-\nu}\sum_{l = 1}^{\infty} \sum_{\mu = 0}^{\infty} (-1)^{\mu} (\mu + \nu)\Psi_l(\varphi_0, \theta_0) \times \\ \times  \iint \limits_{\Omega} \Psi_l(\varphi, \theta) \sin{\theta} C_{\mu}^{\nu}(-\gamma(\varphi, \theta)/\beta) \, d \theta d \varphi \int_0^{+\infty} e^{- \alpha r^2} r^{\frac{3}{2} - \nu}  I_{\lambda_l}(\beta r) I_{\mu + \nu} (\beta r)\,d r. 
\end{multline}
For simplicity we choose $\nu = 1$ and then use the identity (\cite{Ryzik}) for the second integral
\begin{equation}
	\label{I2}
	\begin{aligned}
		I_2^{l, \mu} = \int_0^{\infty} e^{-\alpha r^2} r^{\frac{1}{2}}  I_{\lambda_l}(\beta r) I_{\nu+\mu}(\beta r) d r = 2^{-\lambda_l-\mu-2} \alpha^{-(\lambda_l+\mu+5/2)/2} \beta^{\lambda_l+\mu+1}\\
	\cdot  \frac{\Gamma[(3/2+\mu+\lambda_l)/2]}{\Gamma(\mu+1)\Gamma(\lambda_l+1)} { _3F_3} \left[\begin{array}{ccc}
  \frac{\lambda_l + \mu + 2}{2} & \frac{\lambda_l + \mu + 3}{2} & \frac{\lambda_l + \mu + 5/2}{2} \\ 
  \mu + 2 & \lambda_l + 1 & \mu + \lambda_l + 2 \\
 \end{array}; \frac{\beta^2}{\alpha} \right],
	\end{aligned}
\end{equation}
where $_3F_3(a_1, b_1, c_1; a_2, b_2, c_2; z)$ is a generalized hypergeometric function.

Moreover, with $\nu = 1$ the Gegenbauer polynomials become the Chebyshev polynomials of the second kind, which admit the representation (\cite{Abramowitz})
\begin{equation*}
	U_l(x) = \sum_{k = 0}^{[l/2]} (-1)^k C_k^{l-k} (2x)^{l-2k}, 
\end{equation*}
where $[x]$ is the floor function. Thus, the  integral in \eqref{Q_via_two_int} over $\Omega$ can be represented as

\begin{equation}
	\label{I1}
	\! I_1^{l, \mu} \! = \! \sum_{k=0}^{[\mu/2]} \! 2^{\mu - 2k} (-1)^{\mu-k} C_{k}^{\mu-k} \!\! \iint \limits_{\Omega} \! \Psi_l(\varphi, \theta) \sin{\theta} (\xi_{\alpha} \sin{\theta} \sin \varphi + \xi_{\beta} \sin{\theta} \cos{\varphi} + \xi_{\gamma} \cos{\theta})^{\mu - 2k} d \theta d \varphi.
\end{equation}

As a result, we get the expression for the integral
\begin{equation}
\label{Q_representation}
Q(\tau, r_0, \varphi_0, \theta_0) = 	\frac{2e^{\kappa}}{\beta \tau \sqrt{r_0}} \sum_{l = 1}^{\infty} \sum_{\mu = 0}^{\infty} (-1)^{\mu} (\mu + 1)\Psi_l(\varphi_0, \theta_0) \cdot I_1^{l, \mu} \cdot I_2^{l, \mu}, 	
\end{equation}
where the expression for $I_1^{l, \mu}$ is given in (\ref{I1}) and the expression for $I_2^{l, \mu}$ in (\ref{I2}).

{\color{black}
From a computational point of view, this semi-analytical representation has an advantage in comparison to a direct integration of the Green's function. Consider parameters $M$ and $\mu_{max}$, the limits in the summation \eqref{Q_representation}, and $N$ the number of terms in the eigenvalue expansion of $\Psi_l$. The integrals over $\Omega$ can be computed numerically. Since $\Theta(\theta)$ is smooth, we can transform the domain to a rectangle by a smooth change of variables and use a high order integration method such as Gaussian quadrature. We denote by $N_{\Omega}$ the number of two-dimensional integration nodes. First, we precompute $\Psi_l$ for all $l$ on the grid; it can be done with $O(M N N_{\Omega})$ operations. Then, we precompute the integrals in \eqref{I1}, the complexity of which is is $O(\mu_{max} M N_{\Omega})$. Then, we precompute $I_1^{l, \mu}$, with complexity $O(  \mu_{max}^2 M)$. Finally, we compute \eqref{Q_representation}, using our precomputations, with $O(\mu_{max} M)$ operations. As a result, overall complexity is $O(M N N_{\Omega} +  \mu_{max} M N_{\Omega} + M \mu_{max}^2)$. In comparison, the direct integration of the Green's function with pre-computations of $\Psi_l$ for all $l$  gives $O(M N N_{\Omega} +M N_r N_{\Omega})$ operations, where $N_r$ is the number of quadrature nodes for the variable $r$, and which requires a truncation at some $r_{max}$ in order to deal with the improper integral over $r$.
Note that (\ref{greens_final}) is for the de-drifted Green's function, so that after multiplication by the factor in (\ref{non_drift}) each term in the sum is a non-trivial function of $(r',\phi',\theta')$ and no separation structure can be exploited.
Usually, the values of $\mu_{max}$ and $N$ can be chosen relatively small in comparison to $N_r$ for comparable accuracy.
}

In our computations we use the parameters in Table \ref{table:params_vols}, and liabilities as in Table \ref{table:params_liab}.
\begin{table}[h]
	\begin{center}
		\begin{tabular}{| c | c | c | c | c | c | c | c | c | c | }
			\hline
			 $\sigma_1$ & $\sigma_2$ & $\sigma_3$ & $\rho_{xy}$ & $\rho_{xz}$ & $\rho_{yz} $ & $R_1$ & $R_2$ & $R_3$ & $T$  \\ 
			\hline
			 1 & 1 & 1  & 0.8 & 0.2 & 0.5&  0.4 & 0.45 & 0.4 & 1 \\
			\hline
		\end{tabular}
	\caption{Assumed model parameters.\label{table:params_vols}}		
	\end{center}
\end{table}
\begin{table}[h]
	\begin{center}
		\begin{tabular}{| c | c | c | c | c | c | c | c | c | }
			\hline
			$L_{1}$ & $L_{2}$ & $L_{3}$ & $L_{12}$  & $L_{13}$ & $L_{21}$ &  $L_{23}$ &  $L_{31}$ &  $L_{32}$   \\ 
			\hline
			60 & 70 &  65 & 10 & 15 & 10 & 10 & 5 & 10 \\
			\hline
		\end{tabular}
	\caption{Assumed external and mutual liabilities.\label{table:params_liab}}		
	\end{center}
\end{table}

\begin{figure}[h]
	\begin{center}
				\subfloat[]{\includegraphics[width=0.5\textwidth]{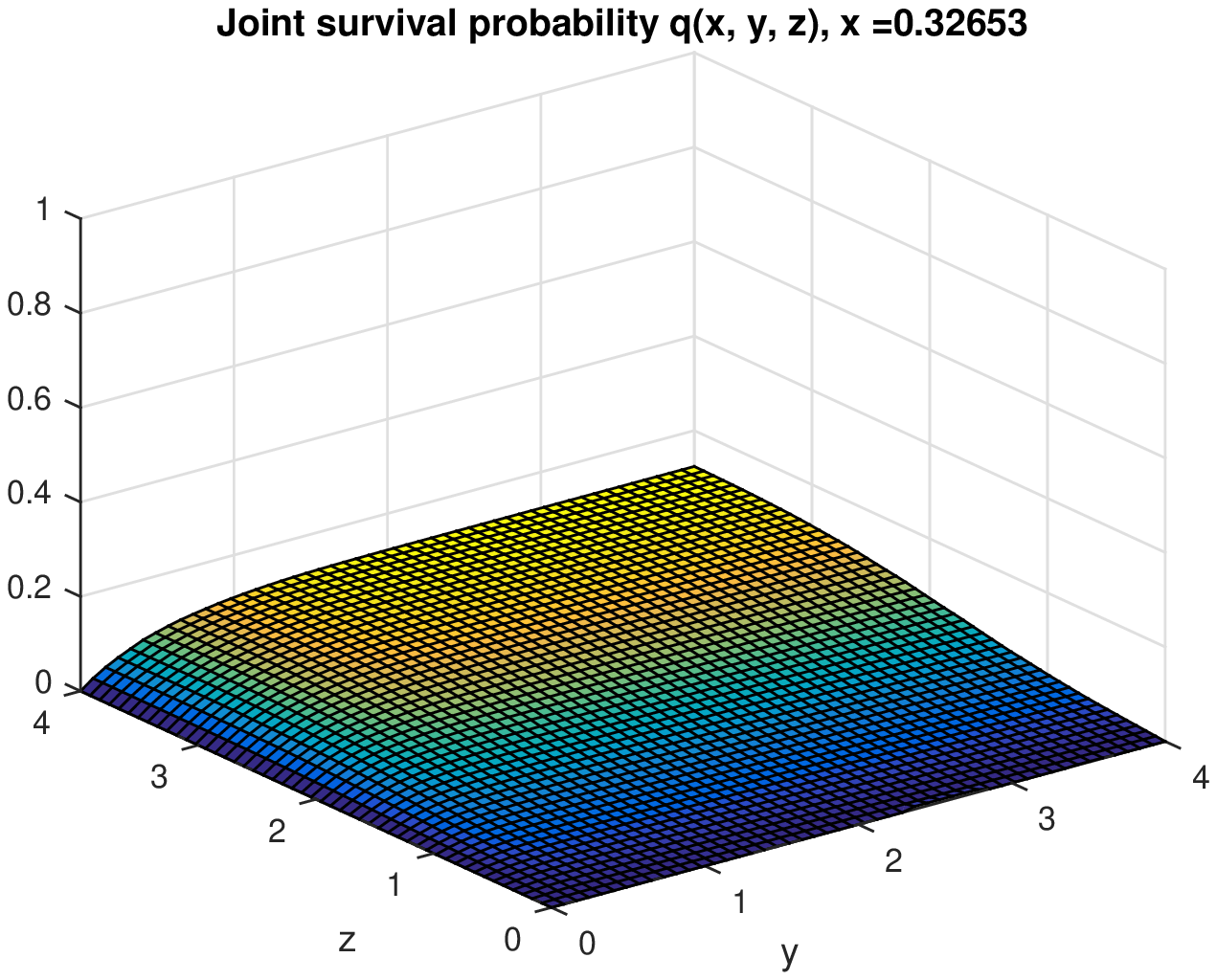}}
				\subfloat[]{\includegraphics[width=0.5\textwidth]{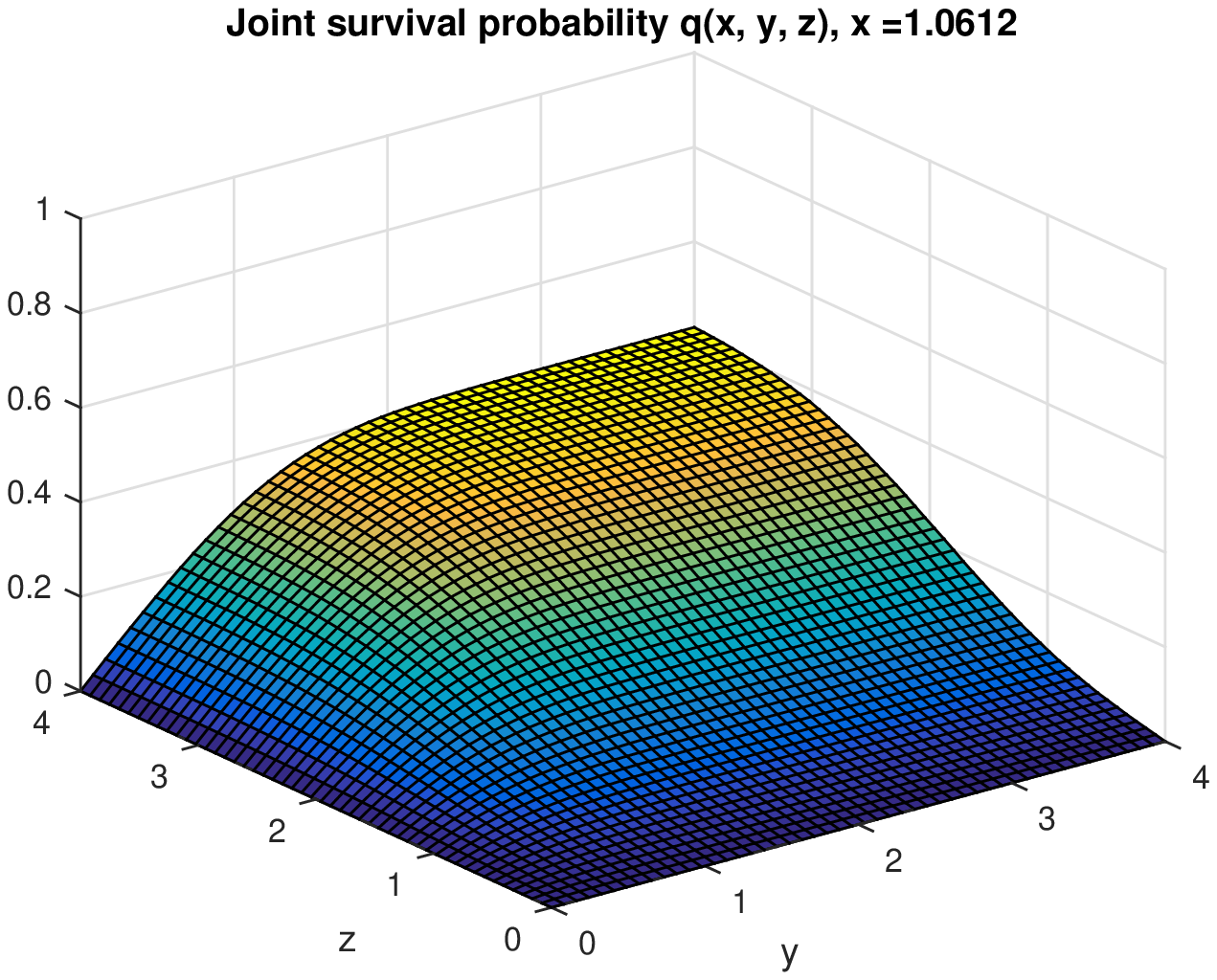}}\\
				\subfloat[]{\includegraphics[width=0.5\textwidth]{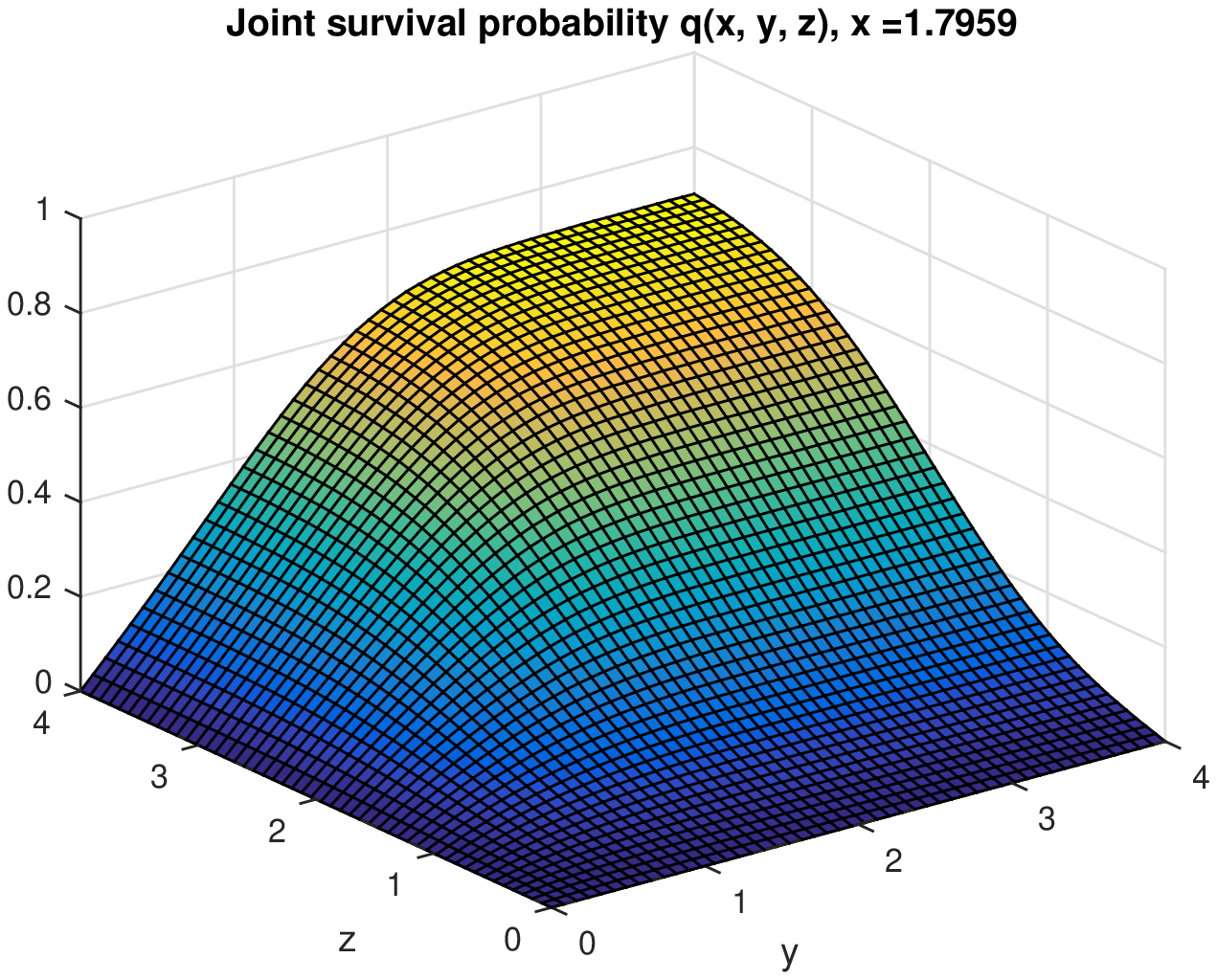}}
				\subfloat[]{\includegraphics[width=0.5\textwidth]{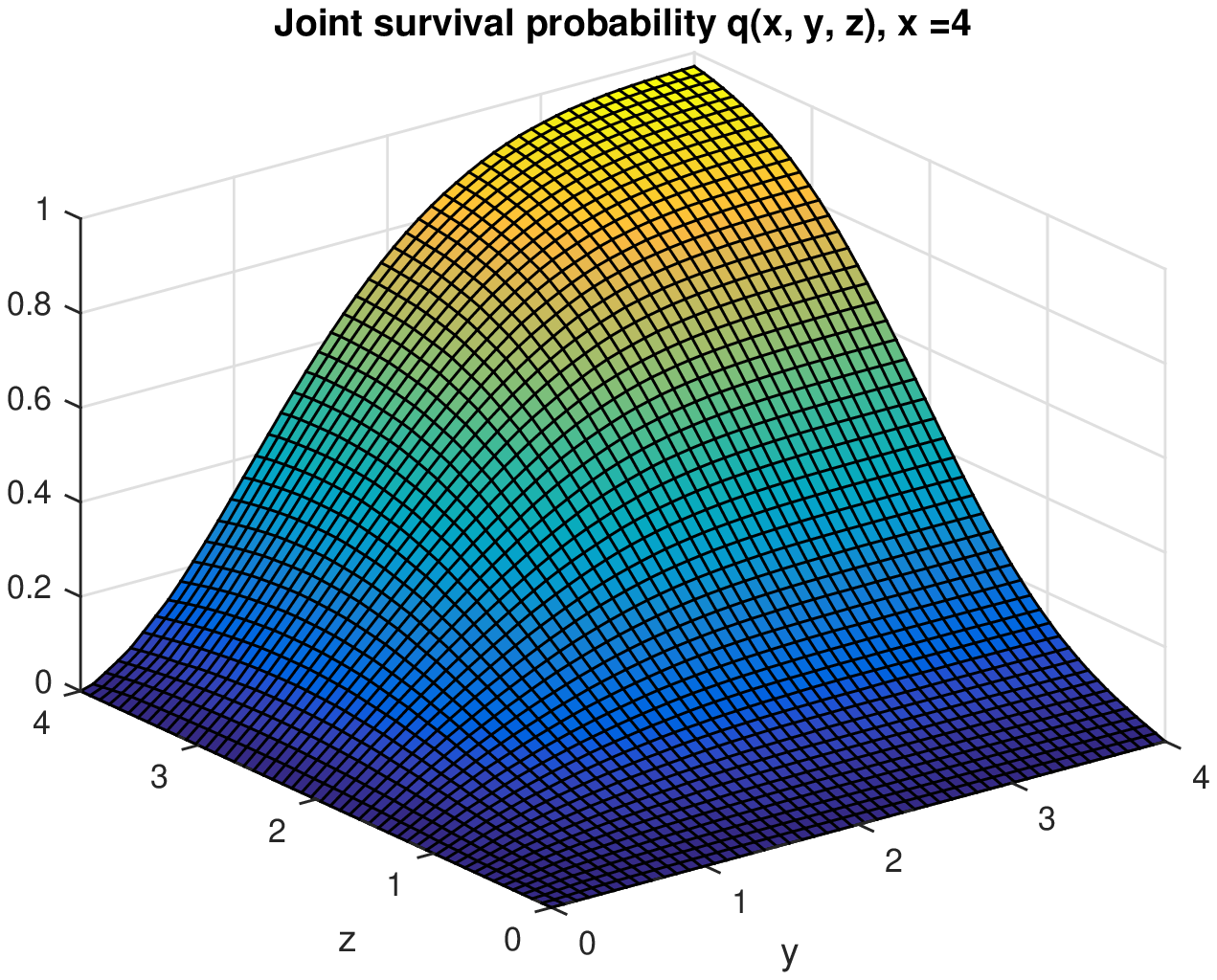}}
	\end{center}		
	\vspace{-20pt}
	\caption{Joint survival probability for 3 banks as a projection on the $y$-$z$ plane for different values of $x$.}
 	\label{joint_prob1}
\end{figure}

In Figure \ref{joint_prob1} we show the computed joint survival probabilities as a projection onto the $y$-$z$ plane for different values of $x$.

\subsection{Computation of Credit and Debt Value Adjustments}
In this section, we compute valuation adjustments for credit default swaps (CDS).
We associate $x$ with the protection seller (PS), $y$ with the protection buyer (PB), and $z$ with the reference name (RN). 

The Credit Value Adjustment (CVA) is the additional cost due to the possibility of the protection seller's default and 
can be written as (see \cite{LiptonSav})
\begin{equation*}
	V^{CVA}(t,x,y,z) = (1 - R_{PS}) \mathbb{E}[\mathbbm{1}_{\tau^{PS} < \min(\tau^{PB}, \tau^{RN}, T)} (V^{CDS}_{\tau_{PS}})_{+}| (X_t,Y_t,Z_t) = (x, y, z)],
\end{equation*} 
where $\tau^{PS}$, $\tau^{PB}$ and $\tau^{RN}$ are the default times of the protection seller, protection buyer and reference name, respectively,
and $V^{CDS}_t$ is the price of a CDS contract on the reference name with non-risky counterparties.

Using Feynman-Kac, we can write the initial-boundary value problem for CVA as
\begin{align*}
		& \frac{\partial V^{CVA}}{\partial t} + \mathcal{L} V^{CVA} = 0, \\
		& V^{CVA}(t, 0, y, z) =  (1 - R_{PS}) V^{CDS}(t, z)_{+}, \\
		& V^{CVA}(t, x, 0, z) = 0, \quad V^{CVA}(t, x, y, 0) = 0, \\
	        & V^{CVA}(T, x, y, z) = 0.
\end{align*}
Then, by an application of Green's formula as in \cite{LiptonSav},
\begin{equation}
	\label{CVA_eq}
        V^{CV\!A}(0, r, \varphi, \theta) =  \frac{1-R_{PS}}{2} \int \limits_0^T \int \limits_0^{\infty} \int \limits_0^{\Theta(0)} \frac{V^{CDS}(t, r', 0, \theta')_{+} G_{\varphi'}(t, r', 0,  \theta', r, \varphi, \theta)}{\sin{\theta'}} \, d \theta' d r' d t, \\
\end{equation}
where $V^{CDS}$ can be found analytically (\cite{LiptonItkin2015}).

Similarly, the Debt Value Adjustment (DVA) represents the additional benefit associated with the default and is given by
\begin{equation*}
	V^{DVA}(t,x,y,z) = (1 - R_{PB}) \mathbb{E}[\mathbbm{1}_{\tau^{PB} < \min(\tau^{PS}, \tau^{RN}, T)} (V^{CDS}_{\tau_{PB}})_{-}| (X_t, Y_t, Z_t) = (x, y, z)].
\end{equation*} 
Then, using again the Feynman-Kac formula
\begin{align*}
		& \frac{\partial V^{DVA}}{\partial t} + \mathcal{L} V^{DVA} = 0, \\
		& V^{DVA}(t, 0, y, z) =  (1 - R_{PS}) V^{CDS}(t, z)_{-}, \\
		& V^{DVA}(t, x, 0, z) = 0, \quad V^{DVA}(t, x, y, 0) = 0, \\
	        & V^{DVA}(T, x, y, z) = 0.
\end{align*}
Thus,
\begin{equation}
	\label{DVA_eq}
        V^{DV\!A}(0, r, \varphi, \theta) =  \frac{R_{PB}-1}{2} \int \limits_0^T \int \limits_0^{\infty} \int \limits_0^{\Theta(0)} \frac{V^{CDS}(t, r', 0, \theta')_{-} G_{\varphi'}(t, r', \bar{\omega},  \theta', r, \varphi, \theta)}{\sin{\theta'}} \, d \theta' d r' d t . \\
\end{equation}

A distinct advantage of our semi-analytical method is that we can compute the derivative $G_{\varphi}$ in \eqref{CVA_eq} and \eqref{DVA_eq} analytically as
\begin{multline*}
		G_{\varphi'}(\tau, r', 0, \theta', r, \varphi, \theta)  =   \kappa (r', 0, \theta', r, \varphi, \theta) \frac{e^{-\frac{r'^2+r^2}{2 \tau}}}{\tau \sqrt{r' r}}\sum_{l = 1}^{\infty} I_{\lambda_l} \left(\frac{r' r}{\tau}\right) \times \\
		\times \left(\sum_{n = 1}^{\infty} c_n^l k_n \tan^{k_n}\left(\frac{\theta'}{2}\right) {}_{2}F_1 \left(\lambda_l, 1 - \lambda_l, 1 + k_n, \sin^2 \frac{\theta'}{2} \right)\right) \times \\
		\times  \left(\sum_{n = 1}^{\infty} c_n^l \sin{\left(k_n \varphi \right)} \tan^{k_n}\left(\frac{\theta}{2}\right) {}_{2}F_1 \left(\lambda_l, 1 - \lambda_l, 1 + k_n, \sin^2 \frac{\theta}{2} \right)\right).
\end{multline*}
Therefore, all terms in \eqref{CVA_eq} and \eqref{DVA_eq} can be computed analytically, and CVA can be computed by integration. As the functions are smooth, this can be done with a higher order integration method, such as Gaussian quadrature.

\begin{figure}[h]
	\begin{center}
				\subfloat[]{\includegraphics[width=0.5\textwidth]{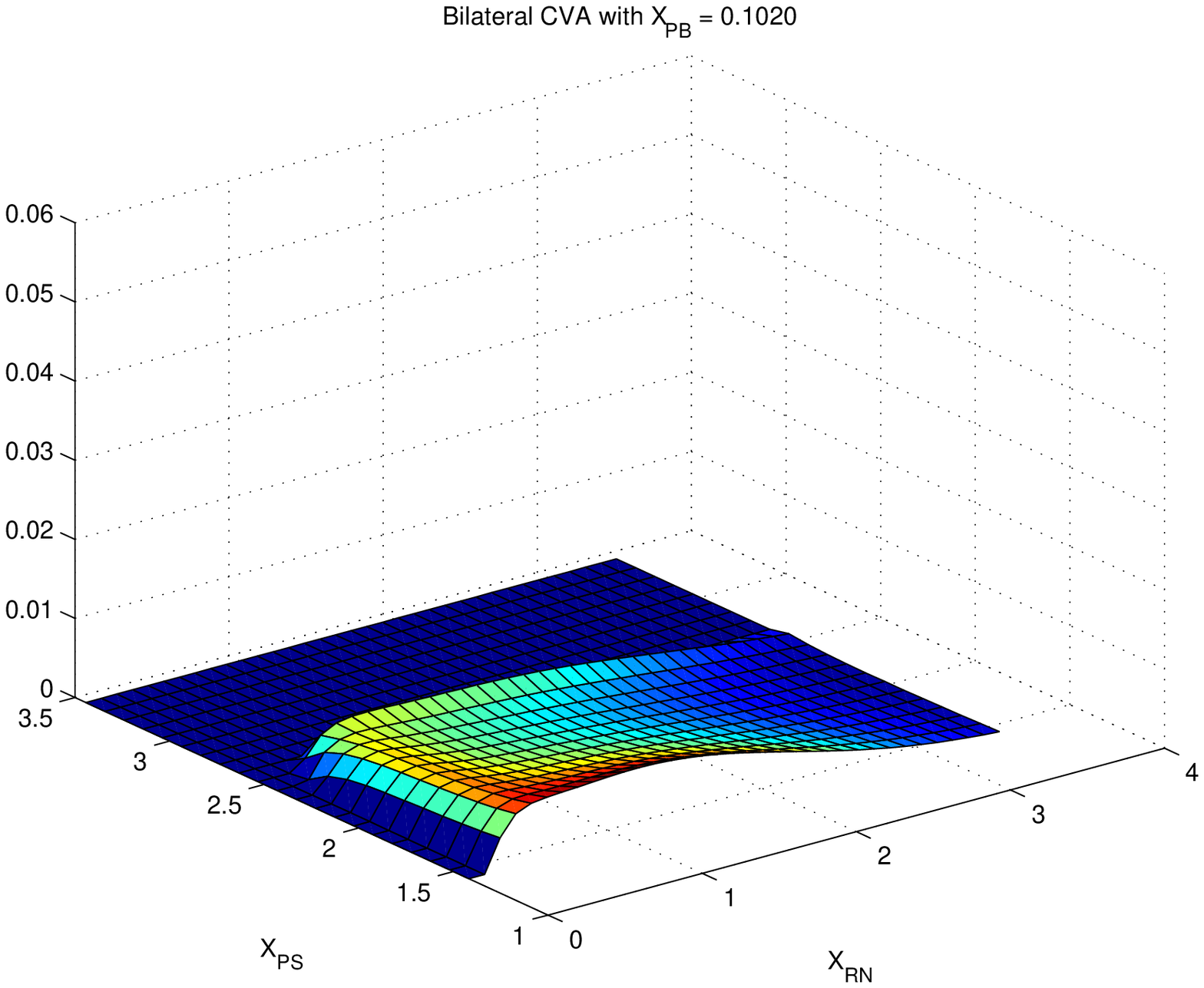}}
				\subfloat[]{\includegraphics[width=0.5\textwidth]{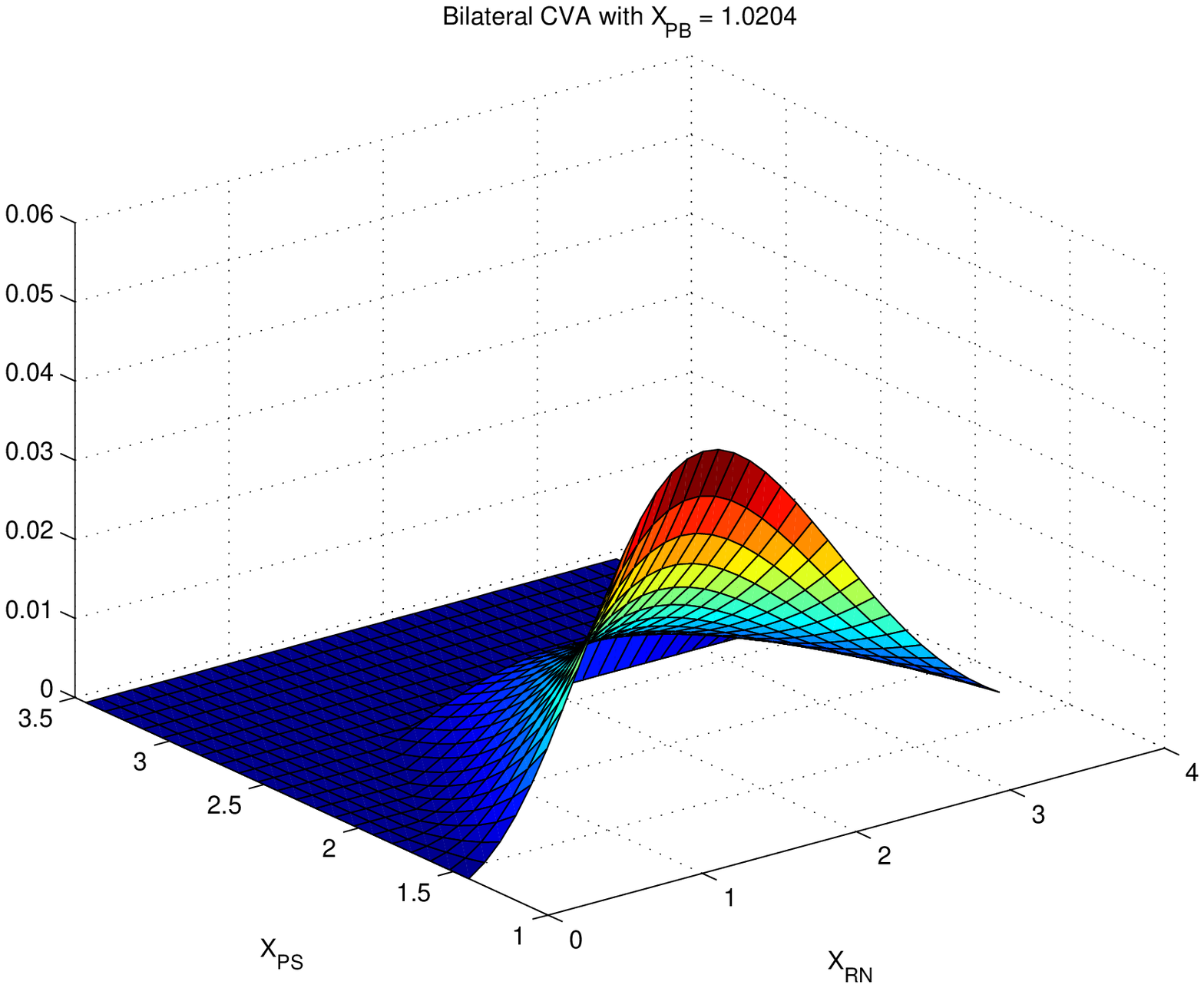}}\\
				\subfloat[]{\includegraphics[width=0.5\textwidth]{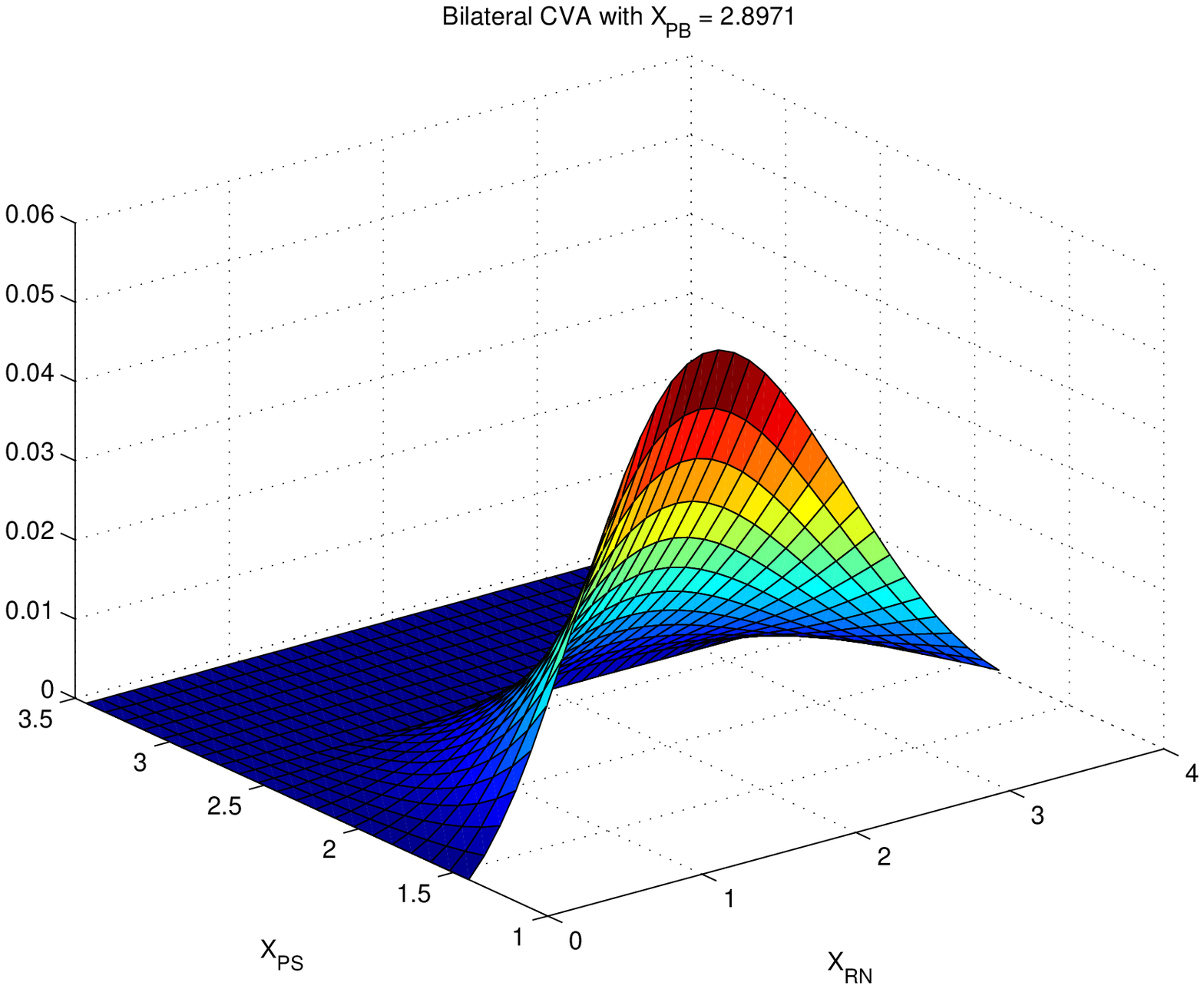}}
				\subfloat[]{\includegraphics[width=0.5\textwidth]{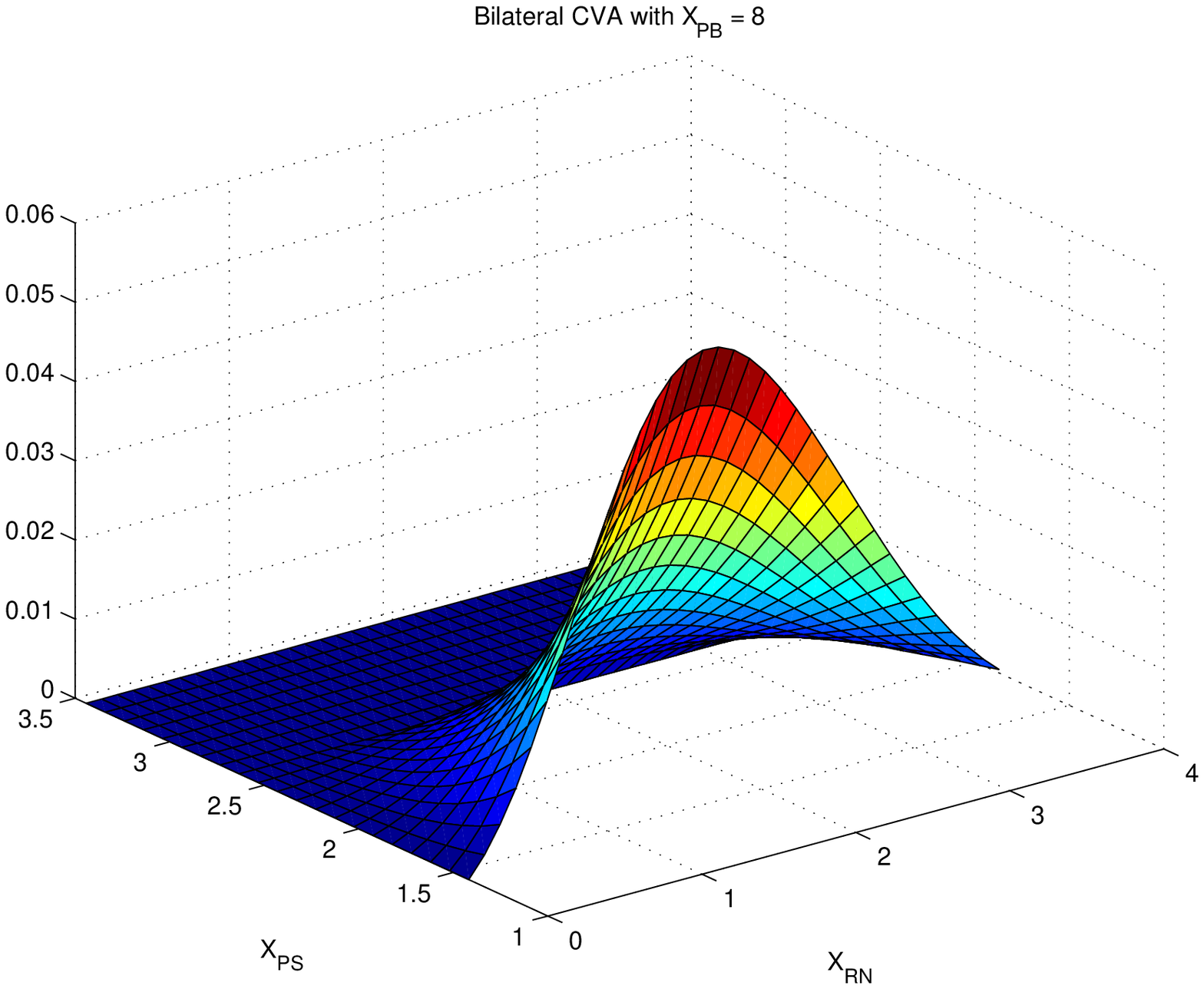}}
	\end{center}		
	\vspace{-20pt}
	\caption{CVA as a projection on the $z$-$x$ plane for different values of $y$. We plot on a truncated domain to better show the impact of the protection buyer on CVA.}
 	\label{cva1}
\end{figure}

In our computations we use the parameters from Tables \ref{table:params_vols} and \ref{table:params_liab}. In Figure \ref{cva1}, we then show the computed CVA as a projection on the $z$-$x$ plane for different values of $y$.

{\color{black}

The CVA decreases with the distance $x$ (or $x_{\rm PS}$) from the default boundary of the protection seller: the less likely the protection seller is to default, the lower the CVA.
In terms of the distance to default $z$ (or $x_{\rm RN}$) of the reference name, the CVA is highest for intermediate values: if the RN is close to default, it is unlikely that the PS defaults before their CDS liability and the CVA is small; if the RN is far from default, it is unlikely that the protection is payable and again the CVA is small. The CVA mostly increases with the value of the protection buyer $y$ (or $x_{\rm PB}$) as their default, which is more likely for small $y$, voids the protection payment. 

For all three firms, the marginal default probability decreases with the (rescaled, logarithmic) firm value. It increases with the firm's volatility. The dependence of the CVA on the marginal default probabilities and volatilities is therefore implicitly given by the plots in Figure \ref{cva1} and we do not show them explicitly for brevity.

The impact of CVA and DVA on CDS prices in this model and its sensitivity to input parameters, specifically and importantly the correlations between the firms, is discussed in more detail in Section 7 of \cite{LiptonSav} (using the finite element approach).
The complexity of the integration remains the same in the finite-element method, but, using our semi-analytical method, we compute derivatives in \eqref{CVA_eq} and \eqref{DVA_eq} analytically.}

\subsection{Comparison with Merton model}

In this section, we compare various important characteristics of the model with a simple Merton-style model as in  \cite{Merton}, where at the terminal time $T$ the settlement process is defined in the spirit of \cite{Eisenberg}. The main difference to the above set-up is that there is no continuous default monitoring as in \cite{BlackCox}.
As a result, the Green's function satisfies the following equation
\begin{eqnarray}
	\label{3d_eq_no_boundary}
			 \frac{\partial G}{\partial t'} &=& \frac{1}{2} G_{x' x'} + \frac{1}{2} G_{y' y'} + \frac{1}{2} G_{z' z'} + \rho_{xy} G_{x' y'} + \rho_{xz} G_{x' z'} + \rho_{yz} G_{y' z'} \\
		&& \hspace{4.5 cm} \nonumber
			- \mu_x G_{x'} - \mu_y G_{y'} - \mu_z G_{z'} , \quad (t', x', y', z') \in \mathbb{R}^4_{+},	
\end{eqnarray}
with the boundary and initial conditions
\begin{equation}
	\label{greens_no_boundary}
	\left\{
		\begin{aligned}
			& G(0, x', y', z', x, y, z) = \delta(x' - x) \delta(y' - y) \delta(z' - z), &\quad (x', y', z') \in \mathbb{R}^3_{+}, \\
			& G(t'-t, x', y', z',  x, y, z) \underset{x'\ \to -\infty}{\longrightarrow} 0, \quad G(t'-t, x', y', z',  x, y, z) \underset{x'\ \to +\infty}{\longrightarrow} 0,& \quad (t', y', z') \in \mathbb{R}^3_{+}, \\
			& G(t'-t, x', y', z',  x, y, z) \underset{y'\ \to -\infty}{\longrightarrow} 0, \quad G(t'-t, x', y', z',  x, y, z) \underset{y'\ \to +\infty}{\longrightarrow} 0,& \quad (t', x', z') \in \mathbb{R}^3_{+},\\
			& G(t'-t, x', y', z',  x, y, z) \underset{z'\ \to -\infty}{\longrightarrow} 0, \quad G(t'-t, x', y', z',  x, y, z) \underset{z'\ \to +\infty}{\longrightarrow} 0,& \quad (t', x', y') \in \mathbb{R}^3_{+},
		\end{aligned}
	\right.
\end{equation}
for given $(t, x, y, z) \in \mathbb{R}^4_{+}$.

The solution of \eqref{3d_eq_no_boundary}, \eqref{greens_no_boundary} can be expressed as
\begin{equation}
	G(t'-t, x', y', z',  x, y, z)  = \frac{1}{\sqrt{(2 \pi)^3 | \Sigma | t}} \exp \left(-\frac{1}{2t} (v' - v - \mu)^T \Sigma^{-1} (v' - v - \mu) \right),
\end{equation}
where $\Sigma$ is the correlation matrix, $\mu = (\mu_x, \mu_y, \mu_z)^T, v = (x, y, z)^T, v' = (x', y', z')^T$.

We compare the results of both models for different sets of correlations.
We calibrate the volatilities for the \cite{Lipton2015} model to individual CDS spreads, while for
the Merton model we use individual bond spreads, because they can be easily evaluated as in the original paper \cite{Merton}.\footnote{For CDSs, one needs to take into account the intermediate payments up to default, but in Merton's model the default can occur only at the terminal time $T$.}
For simplicity, we ignore the dependence (through liabilities) between banks for the calibration step, and calibrate them individually. 

We choose Unicredit,  Santander, and Société Générale as the first, second, and third bank, respectively. The data for external liabilities can be found in banks' balance sheets, which are publicly available. Usually, mutual liabilities data are not public information, thus we make the assumption as in \cite{Kaushansky2017} that they are a fixed proportion of the total liabilities, namely  5\% here. An asset’s value is the sum of liabilities and equity price.
 
 In Table \ref{data_table}, we provide all assets $A_i$ and liabilities $L_i$.
 We take 5-year CDS/bond spread data from Bloomberg. The results of the calibration are given in Table \ref{table:params_1d}.
As expected, a higher volatility is needed in the Merton model than for continuous default monitoring to give the same default probabilities and hence credit spreads.

\begin{table}[H]
	\begin{center}
		\begin{tabular}{| c | c | c | c | c | c |}
			\hline
			 $L_1$ & $A_1$ & $L_2$ & $A_2$ & $L_3$ & $A_3$  \\ 
			\hline
			346.58& 362.96&89.67 &  96.37 & 1607.21 & 1654.38 \\
			\hline
		\end{tabular}
		\caption{Assets and liabilities on 30/06/2017 (Bloomberg).}
		\label{data_table}	
	\end{center}
\end{table}

\begin{table}[H]
	\begin{center}
		\begin{tabular}{| c | c | c | c |}
			\hline
			Model& $\sigma_1$ & $\sigma_2$ & $\sigma_3$  \\ 
			\hline
			 \cite{Lipton2015} & 0.0179 &0.0231& 0.0105\\
			 Merton & 0.0194&  0.0245&   0.0118\\
			\hline
		\end{tabular}
		\caption{Calibrated parameters of one-dimensional models on 30/06/2017 for $T = 5$.}
		\label{table:params_1d}	
	\end{center}
\end{table}

%

Now we consider several sets of parameters for the correlations along with the calibrated volatilities and present the resulting joint survival probabilities in Table \ref{table:results}.
The most striking observation is that while volatilities have been calibrated to give the same marginal survival probabilities, the Merton model now consistently underestimates the joint survival probabilities relative to the continuously monitored model. Had we used the same volatilities for both models, it is clear that the Merton model would have overestimates the joint survival probabilities.

\begin{table}[H]
	\begin{center}
		\begin{tabular}{| c || c | c | c || c | c | c |}
			\hline
			Set& $\rho_{xy}$ & $\rho_{xz}$ & $\rho_{yz}$ & \cite{Lipton2015} & Merton \\ 
			\hline
			1 & 0.0 & 0.0 & 0.0 &  0.6776 & 0.6316 \\
			2  & 0.8&  0.2& 0.5 & 0.7542 & 0.7165 \\
			3 & 0.2 & -0.1 & -0.6 & 0.6739 & 0.6226 \\
			4 & 0.5 & 0.5 & 0.5 & 0.7428 & 0.7060 \\
			5 & 0.1 & -0.1 & -0.2 & 0.6726 & 0.6249 \\
			\hline
		\end{tabular}
		\caption{Joint survival probabilities for different sets of correlation parameters.}
		\label{corr_params}	\label{table:results}	
	\end{center}
\end{table}

\section{Conclusion}

We have found a semi-analytical formula for the three-dimensional transition probability of Brownian motion in the positive octant with absorbing boundaries.
The density is hereby represented as a series of special functions containing unknown parameters, which can be found as the solution of a nonlinear eigenvalue problem.

Applying the results to an extended default model of \cite{Lipton2015} for three banks, we obtained expressions for joint survival probability, CVA and DVA for a CDS. 

This study of the three-dimensional case raised a number of open questions:
(i) Establishing the speed of convergence of the solution as the truncation index $N$ increases, and finding computable error estimates, would be valuable.
(ii) Numerical tests suggest that the nonlinear eigenvalues of $T$ are also nonlinear eigenvalues of $T'$, the derivative of $T$ with respect to the parameter $\lambda$. It would be interesting to investigate this analytically.
(iii) In the uncorrelated case, the $n$-th eigenvalue has multiplicity $n$ (see the last example in Appendix \ref{AppendixComputations}),
while in the tests with non-zero correlations the eigenvalues are fairly evenly spaced with multiplicity 1 (see Table \ref{eigenval_table}). The transition from one case to the other merits further investigation.

{\color{black}
A similar approach may in principle be possible in more than three dimensions.
For an $n$-dimensional transition probability, separation into spherical coordinates, i.e.\ the radius and $n-1$ angles, can be applied as in the three-dimensional case.
In the four-dimensional case specifically,
preliminary computations suggest that a similar representation by special functions can be found.
This involves finding two nested sequences of nonlinear eigenvalues, i.e.\ for each nonlinear eigenvalue as in (\ref{Jc_eq_inf}) a further infinite-dimensional eigenvalue problem has to be solved, in order to match the zero boundary conditions.
The main question here is the complexity and accuracy of the method. We leave this question and the $n$-dimensional problem for future research.
}

\begin{appendices}
\section{Elimination of the cross-derivatives and drift terms}
\label{cross_deriv}
The correlation matrix of the Brownian motion is positive definite and has the form
\begin{equation*}
	\Sigma = \left[
		\begin{matrix}
			1 & \rho_{xy} & \rho_{xz} \\
			\rho_{xy} & 1 & \rho_{yz} \\
			\rho_{xz} & \rho_{yz} & 1
		\end{matrix}
	\right] > 0.
\end{equation*}
Then equation (\ref{adjoint_eq}) can be rewritten as
\begin{multline}
	\label{3d_eq}
			 \frac{\partial G}{\partial t'} = \frac{1}{2} G_{x' x'} + \frac{1}{2} G_{y' y'} + \frac{1}{2} G_{z' z'} + \rho_{xy} G_{x' y'} + \rho_{xz} G_{x' z'} + \rho_{yz} G_{y' z'} \\
			- \mu_x G_{x'} - \mu_y G_{y'} - \mu_z G_{z'} , \quad (t', x', y', z') \in \mathbb{R}^4_{+}, \\	
\end{multline}
with the boundary and initial conditions
\begin{equation*}
	\left\{
		\begin{aligned}
			& G(0, x', y', z', x, y, z) = \delta(x' - x) \delta(y' - y) \delta(z' - z), &\quad (x', y', z') \in \mathbb{R}^3_{+}, \\
			& G(t'-t, 0, y', z', x, y, z) = 0, \quad G(t'-t, x', y', z',  x, y, z) \underset{x'\ \to +\infty}{\longrightarrow} 0,& \quad (t', y', z') \in \mathbb{R}^3_{+}, \\
			& G(t'-t, x', 0, z', x, y, z) = 0, \quad G(t'-t, x', y', z',  x, y, z) \underset{y'\ \to +\infty}{\longrightarrow} 0,& \quad (t', x', z') \in \mathbb{R}^3_{+},\\
			& G(t'-t, x', y', 0, x, y, z) = 0, \quad G(t'-t, x', y', z',  x, y, z) \underset{z'\ \to +\infty}{\longrightarrow} 0,& \quad (t', x', y') \in \mathbb{R}^3_{+},
		\end{aligned}
	\right.
\end{equation*}
for given $(t, x, y, z) \in \mathbb{R}^4_{+}$.

To eliminate the cross-derivatives, we introduce the following change of variables (\cite{Lipton2014}, \cite{LiptonSav})
\begin{equation*}
	\left\{
	\begin{aligned}
		& \alpha(x, y, z) = x, \\
		& \beta(x, y, z) = - \frac{1}{\bar{\rho}_{xy}} (\rho_{xy} x - y), \\
		& \gamma(x, y, z) = \frac{1}{\bar{\rho}_{xy} \chi} \left[(\rho_{xy} \rho_{yz} - \rho_{xz})x + (\rho_{xy} \rho_{xz} - \rho_{yz})y + \bar{\rho}_{xy}^2 z \right],
	\end{aligned}
	\right.
\end{equation*}
where $\bar{\rho}_{xy} = \sqrt{1 - \rho_{xy}^2}$ and $\chi = \sqrt{1 - \rho_{xy}^2 - \rho_{xz}^2 - \rho_{yz}^2 + 2\rho_{xy} \rho_{xz} \rho_{yz}}$.  It transforms the equation (\ref{3d_eq}) to
\begin{equation}
	\label{3d_eq_non_corr}
	\left\{
		\begin{aligned}
			& \frac{\partial G}{\partial t'} = \frac{1}{2} G_{\alpha' \alpha'} + \frac{1}{2} G_{\beta' \beta'} + \frac{1}{2} G_{\gamma' \gamma'} - \mu_{\alpha} G_{\alpha'} - \mu_{\beta} G_{\beta'} - \mu_{\gamma} G_{\gamma'}, \\
			& G(0, \alpha', \beta', \gamma', \alpha, \beta, \gamma) = \delta(\alpha' - \alpha) \delta(\beta' - \beta)  \delta(\gamma' - \gamma), 
		\end{aligned}
	\right.
\end{equation}
with zero boundary conditions on the new domain
\begin{equation*}
	D = \{\omega_1 e_1 + \omega_2 e_2 + \omega_3 e_3 | \omega_i > 0 \},
\end{equation*}
where 
\begin{equation*}
		e_1 = (0, 0, 1), \quad e_2 = \left(0, \frac{\chi}{\bar{\rho}_{xy} \bar{\rho}_{xz}}, - \frac{\rho_{yz} - \rho_{xz} \rho_{xy}}{\bar{\rho}_{xy} \bar{\rho}_{xz}} \right), \quad e_3 = \left(\frac{\chi}{\bar{\rho}_{yz}}, -\frac{\rho_{xy} \chi}{\bar{\rho}_{xy} \bar{\rho}_{yz}}, - \frac{\rho_{xz} - \rho_{yz} \rho_{xy}}{\bar{\rho}_{xy} \bar{\rho}_{yz}} \right),
\end{equation*}
and where
\begin{align*}
		& \mu_{\alpha} = \mu_{x}, \\
		& \mu_{\beta} = - \frac{1}{\bar{\rho}_{xy}} (\rho_{xy} \mu_x - \mu_y), \\
		& \mu_{\gamma} = \frac{1}{\bar{\rho}_{xy} \chi} \left[(\rho_{xy} \rho_{yz} - \rho_{xz}) \mu_x + (\rho_{xy} \rho_{xz} - \rho_{yz}) \mu_y + \bar{\rho}_{xy}^2 \mu_z \right].
\end{align*} 

Now we can eliminate the drift term by considering $\tilde{G}(t'-t, x', y', z', x, y, z)$ from (\ref{non_drift}),
which satisfies the following PDE 
\begin{equation*}
	\label{3d_eq_no_drift}
	\left\{
		\begin{aligned}
			& \frac{\partial \tilde{G}}{\partial t'} = \frac{1}{2} \tilde{G}_{\alpha' \alpha'} + \frac{1}{2} \tilde{G}_{\beta' \beta'} + \frac{1}{2} \tilde{G}_{\gamma' \gamma'}, \\
			& \tilde{G}(0, \alpha', \beta', \gamma', \alpha, \beta, \gamma) = \delta(\alpha' - \alpha) \delta(\beta' - \beta)  \delta(\gamma' - \gamma),
		\end{aligned}
	\right.
\end{equation*}
with zero boundary conditions on $\partial D$.

We omit tilde for convenience in the following. In order to take advantage of the wedge-shape of the domain, we change the variables to spherical coordinates

\begin{equation*}
	\left\{
		\begin{aligned}
			& \alpha = r \sin{\theta} \sin{\varphi}, \\
			& \beta = r \sin{\theta} \cos{\varphi}, \\
			& \gamma = r \cos{\theta},
		\end{aligned}
	\right.
\end{equation*}
where $r$ varies from 0 to $+\infty$. In order to determine the range of values of $\varphi$, we project $e_1$ and $e_2$ onto the plane $(\alpha, \beta)$ and get normalized vectors
\begin{equation*}
	\begin{aligned}
		& H_1 = (0, 1), \\
		& H_2 = (\bar{\rho}_{xy}, \rho_{xy}).
	\end{aligned}
\end{equation*}

Thus, $\varphi$ varies from $0$ to 
\begin{equation*}
	\bar{\omega} = \arccos(-\rho_{xy}),
 \end{equation*}
and $\theta$ varies from $0$ to $\Theta(\varphi)$, where $\Theta(\varphi)$ can be expressed in parametric form as in (\ref{varphi}) and (\ref{theta}) (see \cite{LiptonSav} for details).

Now we can rewrite the equation (\ref{3d_eq_non_corr}) in spherical coordinates, as in (\ref{3d_eq_spherical}).

  \section{Schrödinger equation with Pöschl–Teller potential} 
  \label{AppendixComputations}
  Consider the Schrödinger equation with Pöschl–Teller potential on a half-line,
\begin{equation}
\label{poeschlteller}
		 \Upsilon'' - \left(k^2 - \frac{\lambda (\lambda - 1)}{\cosh^2(\zeta)}  \right) \Upsilon = 0
\end{equation}
and boundary condition
\begin{equation}
	\label{inf_cond}
	 \Upsilon(-\infty) = 0.
\end{equation}

\subsection{Fundamental solution}

We follow \cite{Flugge} for the solution of (\ref{poeschlteller}).
Consider the change of variables $y = \cosh^2(\zeta)$. Then, (\ref{poeschlteller}) can be rewritten as
\begin{equation*}
	y (1-y) \Upsilon'' + \left(\frac{1}{2} - y\right) \Upsilon' + \left(\frac{k^2}{4} - \frac{\lambda (\lambda - 1)}{4 y} \right) \Upsilon = 0.
\end{equation*}
Then, if we consider $v(y) = y^{-\lambda/2} \Upsilon(y)$, we get the hypergeometric equation
\begin{equation*}
	y (1-y) v'' + \left(\left(\lambda + \frac{1}{2}\right) - (\lambda + 1) y \right) v' + \frac{1}{4} \left(k^2- \lambda^2 \right) v = 0.
\end{equation*}
The solution of this equation is
\begin{equation*}
	v(y) =  A \cdot {}_{2}F_1 \left(a, b, \frac{1}{2}, 1-y \right) + B \sqrt{1-y} \cdot {}_{2}F_1 \left(a + \frac{1}{2}, b + \frac{1}{2}, \frac{3}{2}, 1 - y \right),
\end{equation*}
where
\begin{equation*}
	a = \frac{1}{2} \left(\lambda - k\right), \quad b = \frac{1}{2} \left(\lambda + k\right).
\end{equation*}
As a fundamental system we consider standard odd and even real solutions:
\begin{align*}
	& \Upsilon_e(\zeta) = \cosh^{\lambda}(\zeta)  {}_{2}F_1 \left(a, b, \frac{1}{2}, -\sinh^2(\zeta) \right), \\
	& \Upsilon_o(\zeta) = \cosh^{\lambda}(\zeta)  \sinh(\zeta) {}_{2}F_1 \left(a + \frac{1}{2}, b + \frac{1}{2}, \frac{3}{2}, -\sinh^2(\zeta) \right),
\end{align*}
where ${}_2F_1(a, b, c, z)$ is the Gaussian hypergeometric function.

Any solution can be represented as a linear combination of the fundamental system,
\begin{equation}
	\label{ups_eq}
	\Upsilon(\zeta) = A \Upsilon_e(\zeta) + B \Upsilon_o(\zeta).
\end{equation}

Now consider an asymptotic  expansion for large $\zeta$. Unlike \cite{Flugge}, who consider the asymptotics at $\pm \infty$ (see the case $Z \equiv \infty$ below), if $Z(\varphi)$ is bounded above, we only consider the asymptotics at $-\infty$.

\subsection{General boundary conditions}

From the known expansion of the hypergeometric function we get
\begin{equation*}
	 \Upsilon_e(\zeta) \to 2^{-\lambda} e^{\lambda |\zeta|} \Gamma \left(\frac{1}{2}\right)  \left[\frac{\Gamma(b-a)}{\Gamma(b) \Gamma \left(\frac{1}{2} - a\right)} 2^{2a} e^{-2a |\zeta|}  + \frac{\Gamma(a-b)}{\Gamma(a) \Gamma \left(\frac{1}{2} - b\right)} 2^{2b} e^{-2b |\zeta|} \right] \label{asympt_exp1}
\end{equation*}
and
\begin{multline*}
	 \Upsilon_o(\zeta) \to -2^{-\lambda-1} e^{(\lambda+1) |\zeta|} \Gamma \left(\frac{3}{2}\right)  \left[\frac{\Gamma(b-a)}{\Gamma \left(b + \frac{1}{2} \right) \Gamma \left(1- a\right)} 2^{2a+1} e^{-(2a+1) |\zeta|}  + \right. \\
	\left. + \frac{\Gamma(a-b)}{\Gamma \left(a + \frac{1}{2} \right) \Gamma \left(1 - b\right)} 2^{2b+1} e^{-(2b+1) |\zeta|} \right]. \label{asympt_exp2}
\end{multline*}

Assume $k > 0$. First, consider the case $\lambda - k = 2l$. Thus, $\Gamma \left(\frac{1}{2} - a \right)$ is singular and  $\Upsilon_e(\zeta)$ satisfies the condition (\ref{inf_cond}). Similar, if $\lambda - k = 2l + 1$, then $\Gamma \left(1-a\right)$ is singular and $\Upsilon_o(\zeta)$ satisfies the condition (\ref{inf_cond}).  

Hence, in order to satisfy the condition (\ref{inf_cond}) we can choose
\begin{equation*}
	A = \begin{cases}
	 \Gamma(b) \Gamma \left(\frac{1}{2} - a \right),  & \lambda - k \notin \mathbb{Z}_+, \\
	 1, & \lambda - k = 2 l\\
	 0, & \lambda - k = 2 l + 1\\
	 \end{cases}, B = \begin{cases}
	 	2 \Gamma \left(b+\frac{1}{2} \right) \Gamma \left(1-a\right) , & \lambda - k \notin \mathbb{Z}_+, \\
		0,&  \lambda - k = 2l, \\
		1, & \lambda - k = 2l + 1.
	\end{cases}
\end{equation*}

Using identities for hypergeometric functions we can simplify (\ref{ups_eq}) to
\begin{equation*}
	\Upsilon(\zeta) = c \cdot e^{k \zeta} {}_{2}F_1 \left(\lambda, 1 - \lambda, 1 + k, \frac{e^{2 \zeta}}{1+e^{2 \zeta}} \right),
\end{equation*}
for all $\lambda$ and $k$.

\subsection{Special case $\rho_{xy} = \rho_{xz} = 0$, $\rho_{yz} = 1$, i.e.\ $Z \equiv \infty$}
Consider first the case when $Z(\varphi)$ is unbounded. 
Now we have to take into account also the asymptotics at $+\infty$. We note that $\Upsilon_e(-\zeta) = \Upsilon_e(\zeta)$ and $\Upsilon_o(-\zeta) = -\Upsilon_o(\zeta)$. The first term in $\Upsilon_o(\zeta)$ and $\Upsilon_e(\zeta)$ goes to infinity unless $\Gamma(\frac{1}{2} - a)$ and $\Gamma(1-a)$ are singular, respectively. The gamma function is singular if the argument is a negative integer. Thus,
\begin{equation*}
	\lambda_l^r = \frac{\pi r}{\bar{\omega}} + l
\end{equation*}
and
\begin{equation*}
	\Psi_l^r(\varphi, \zeta) =
	\begin{cases}
		c \sin(k_r \varphi) \Upsilon_e(\zeta), l \text{ is even}, \\
		c \sin(k_r \varphi) \Upsilon_o(\zeta), l \text{ is odd}.
	\end{cases} 
\end{equation*}

\subsection{Special case $\rho_{xy} = \rho_{xz} = \rho_{yz} = 0$, i.e.\ $Z \equiv 0$}
\label{app:zerocorr}
In this case $\bar{\omega} = \frac{\pi}{2}$.
We can easily see that $\Upsilon_e(0) = 1$ and $\Upsilon_o(0) = 0$ for any $k$ and $\lambda$. Thus, $A = 0, B = 1$. Now we want to find the corresponding eigenvalues. Applying boundary condition (\ref{inf_cond}), we get
\begin{equation*}
	\lambda_l = 2l + 1, n \ge 0,
\end{equation*}
and the corresponding eigenvectors are
\begin{equation*}
	\Psi_l^r(\varphi, \zeta) = c \sin(k_r \varphi) \Upsilon_o(\zeta),
\end{equation*}
where $1 \le r \le l $.

\end{appendices}

\bibliographystyle{apalike}
\bibliography{main}

\end{document}